\documentclass[journal]{IEEEtran}

\usepackage[authoryear,round]{natbib}
\usepackage{amsmath,amssymb,amsfonts}
\usepackage{algorithmic}
\usepackage{graphicx}
\usepackage{textcomp}
\usepackage{xcolor}
\usepackage{booktabs} 
\usepackage{gensymb}
\usepackage{array}
\usepackage{chemist}
\usepackage{tcolorbox}
\usepackage{float}
\usepackage{hyperref}
\usepackage[margin=0.8in]{geometry}
\usepackage{lscape}
\usepackage{adjustbox}
\usepackage{indentfirst}
\usepackage{multicol, blindtext}
\usepackage{pifont}
\usepackage{lettrine}
\usepackage{pdfpages}
\usepackage{colortbl}
\usepackage{lipsum}
\usepackage{colortbl}
\usepackage{lipsum}
\usepackage{threeparttable}
\usepackage{multirow}
\usepackage{longtable}
\usepackage{threeparttablex}
\usepackage{ragged2e}
\usepackage{tabularx}
\usepackage{soul}
\usepackage{tikz}
\usepackage{cancel}

\PassOptionsToPackage{hyphens}{url}
\usepackage{hyperref}

\usepackage{ragged2e}
\usepackage{verbatim}

\makeatletter

\renewcommand{\subsubsection}{%
  \@startsection{subsubsection}{3}{\z@}%
    {1.2ex plus .2ex minus .2ex}%
    {0.5ex plus .2ex}%
    {\normalfont\small\itshape}%
}

\makeatother

\makeatother

\newcommand{\coo}{CO\textsubscript{2}}
\newcommand{\cooeq}[1]{CO\textsubscript{2}eq{#1}}

\newcommand{\addref}[1]{\textcolor{green}{[REF]}}

\newcommand{\No}{\textcolor{red}{\ding{55}}}
\newcommand{\Yes}{\ding{51}}

\soulregister\ref7
\soulregister\pageref7
\soulregister\eqref7
\soulregister\autoref7
\soulregister\cref7
\soulregister\Cref7
\soulregister\cite7
\soulregister\citep7
\soulregister\citet7
\soulregister\appendixCaseStudy0
\soulregister\appendixMethods0
\soulregister\appendixUncertainty0
\soulregister\appendixSupplementary0

\setstcolor{red}

\definecolor{forestgreen}{rgb}{0.13, 0.55, 0.13}%

\newcommand{\strikereview}[1]{\ifbool{printstrike}{\textcolor{blue}{\st{#1}}}{}}

\newcommand{\appendixComparison}{Appendix~A}
\newcommand{\appendixEM}{Appendix~B}
\newcommand{\appendixCC}{Appendix~C}
\newcommand{\appendixCaseStudy}{Appendix~D}

\newbool{printstrike}
\boolfalse{printstrike}

\hypersetup{%
colorlinks=true,
linkcolor=black,
citecolor=black,
urlcolor=blue}

\begin{document}
\raggedbottom
\setcounter{page}{1}
\title{
UNCASExt -- A Systematic Computational Framework for Uncertainty Propagation and Scope Consistency in Absolute Environmental Sustainability Assessments (AESA)}

\author{
\large{Erwan Ike de Bantel$^{a,*}$, Thibault Pirson$^{b}$, Gonzalo Puig-Samper$^{c}$, Jan Marcus Hartmann$^{d}$, \par David Bol$^{b}$, Ghada Bouillass$^{a}$, Bernard Yannou$^{a}$, Marija Jankovic$^{a}$, Michael Hauschild$^{e,f}$} \\
\footnotesize{
\textit{$^{a}$ Université Paris-Saclay, CentraleSupélec, Laboratoire Genie Industriel, 91190 Gif-sur-Yvette, France}~\\
\textit{$^{b}$ Université catholique de Louvain, 1348 Louvain-la-Neuve, Belgium}~\\
\textit{$^c$ Luxembourg Institute of Science and Technology, 5 Avenue des Hauts-Fourneaux, Esch-sur-Alzette, L-4362, Luxembourg }~\\
\textit{$^d$ Institute of Technical Thermodynamics, RWTH Aachen University, Schinkelstraße 8, 52062 Aachen, Germany}~\\
\textit{$^{e}$ Section for Quantitative Sustainability Assessment, Department of Environmental and Resource Engineering, Technical University of Denmark, 2800 Kgs. Lyngby, Denmark}~\\
\textit{$^{f}$ Centre for Absolute Sustainability, Technical University of Denmark, 2800 Kgs. Lyngby, Denmark
}~\\
  \vspace{0.5em}
$^*$\textit{Corresponding author: erwan.de-bantel@centralesupelec.fr.}
}
}

\maketitle
\thispagestyle{empty}

\begin{abstract} 

Absolute environmental sustainability assessment~(AESA) has gained increasing attention in environmental research and policymaking. However, its reliability is challenged by several sources of uncertainty that remain insufficiently accounted for, as well as by scope inconsistencies within the absolute sustainability ratio~(ASR), which compares estimated environmental burdens with allocated carrying capacities for a given human activity. This work introduces \textit{UNCASExt}, an extension of the UNCASE framework for systematically propagating uncertainty and ensuring scope consistency in AESA, together with a supporting open-source Python package, \texttt{pyaesa}. At country and sector levels, the computational framework formalizes allocation procedures that match the scope of allocated carrying capacities with that of estimated environmental burdens across three dimensions: impact pathway modeling; production-based versus consumption-based accounting; business-to-consumer~(B2C) versus business-to-business~(B2B) activities. It also incorporates temporal dynamics, supporting both retrospective and prospective assessments with either static steady-state or dynamic carrying capacities, including greenhouse gas budgets from the Intergovernmental Panel on Climate Change Sixth Assessment Report under Shared Socioeconomic Pathway transition scenarios. The framework is applied to a case study of electricity consumption in France over the period 2019--2060. The results show that mismatches between the functional units of estimated environmental burdens and allocated carrying capacities can lead to substantial underallocation, with a median factor of 4.6$\times$ across all available sector-region pairs in the multi-regional input-output table EXIOBASE 3.10.2. Overall, \textit{UNCASExt} and \texttt{pyaesa} provide a scalable solution to support AESA harmonization and a versatile way forward to bridge the gap between methodological guidelines and practical application.

\end{abstract}


\section*{Graphical abstract}

\begin{figure}[H]
    \centering
    \includegraphics[width=1\columnwidth]{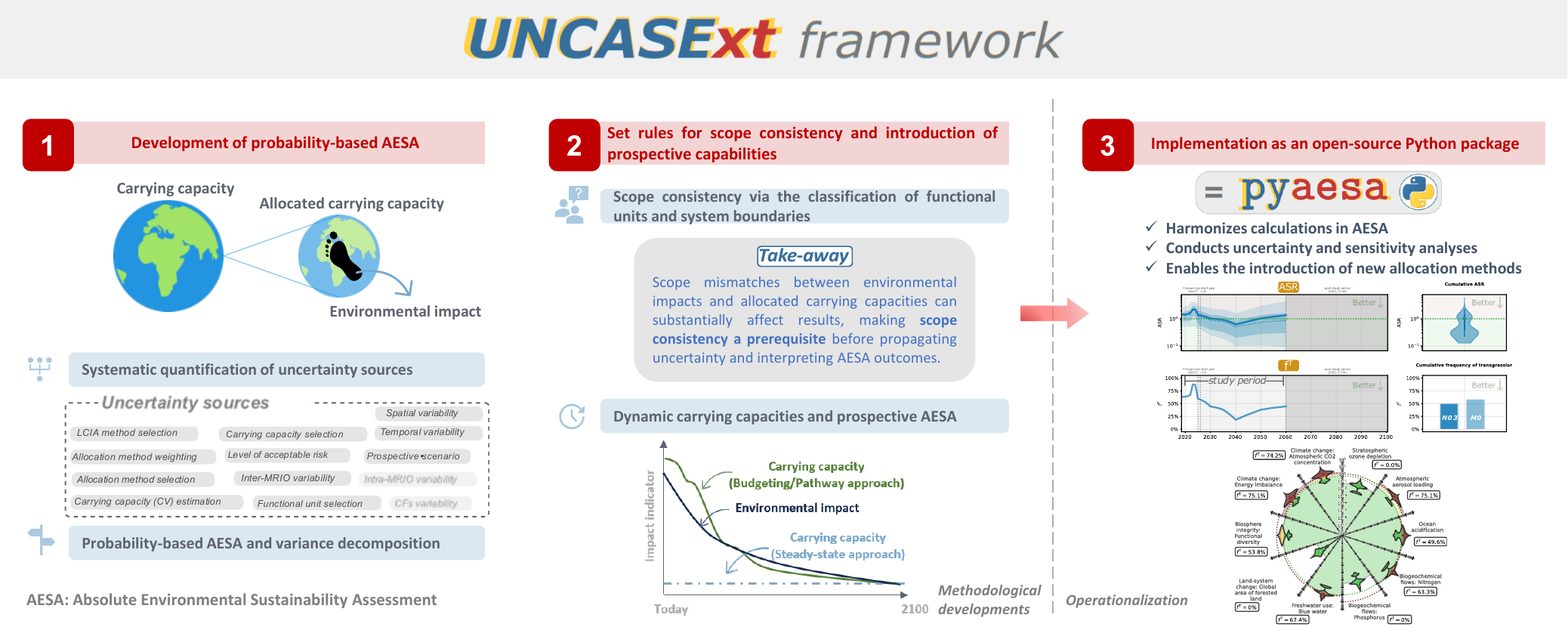}
    \label{fig:fig-intro}
\end{figure}

\section{Introduction}

Absolute environmental sustainability assessment~(AESA) aims to introduce a paradigm shift in environmental assessment by moving beyond relative assessment towards evaluating activities against “absolute” environmental thresholds~\citep{bjorn2020}. These thresholds, referred to as carrying capacities (CCs), represent the maximum environmental burdens that natural systems can sustain without triggering structural or functional changes to the Earth system that are irreversible or difficult to restore \citep{bjorn2015introducing}. Carrying capacities can be allocated to human activities according to normative choices grounded in different distributive justice theories~\citep{hayha2016fairshares, ryberg2020}. Because these activities can be defined at different scales, including countries, sectors, companies, goods and services, and persons, AESA applications increasingly involve diverse allocation approaches and system boundaries. This methodological diversity is becoming increasingly important as AESA gains attention within the environmental assessment research community~\citep{bjornguidanceJRC} and is progressively adopted beyond academia to inform policymaking.

Notable applications include assessments conducted at the European Union level \citep{europeanenvironmentagency2020} and at national scales, such as in the Netherlands, Taiwan, New Zealand, or Australia \citep{lind2024}. At the corporate level, the most widely adopted approach to absolute environmental sustainability is the Science-Based Targets initiative (SBTi) \citep{sbti2025}, which enables companies to set and report science-based greenhouse gas (GHG) emission reduction targets. \par 

While AESA provides a structured approach to benchmark human activities against environmental thresholds, its results are inherently subject to uncertainty coming from incomplete knowledge of input data, normative choices of allocation methods, modeling limitations, etc.~\citep{hjalsted2021, ryberg2021}. To systematically account for uncertainty in AESA studies, a first framework was proposed by \citep{puig2025quantifying}: the UNCASE framework (UNCertainty assessment of Absolute environmental Sustainability Evaluation). It formalizes how uncertainties should be identified, parameterized, and jointly propagated through Monte Carlo simulations. 
Nevertheless, several methodological gaps remain, including insufficient attribution of result variability to individual uncertainty sources and scope inconsistencies between allocated carrying capacities and estimated environmental burdens across impact pathway modeling, production-based versus consumption-based accounting, and business-to-consumer~(B2C) versus business-to-business~(B2B) activities. In addition, the lack of software for reproducible, systematic implementation can hinder broader adoption within the AESA community, especially given the large number of uncertainty sources to be considered and integrated in practice.  \par 

\begin{table}[t!]
\centering

{\fontsize{8}{9.5}\selectfont
\renewcommand{\arraystretch}{0.9}

\resizebox{1\columnwidth}{!}{%

\begin{tabular}{
    p{0.18\columnwidth} p{0.73\columnwidth}}
\toprule[2pt]

\textbf{Abbreviations} & \textbf{Description} \\
\toprule[2pt]

aCC & Allocated carrying capacity \\
AESA & Absolute environmental sustainability assessment \\
aSoCC & Allocated share of carrying capacity \\
ASR & Absolute sustainability ratio \\
CBA & Consumption-based accounting \\
CC & Carrying capacity \\
CoV & Coefficients of variation \\
MRIO & Multi-regional input-output \\
EE-MRIO & Environmentally-extended MRIO \\
$f^T$ & Frequency of transgression \\
GDP & Gross domestic product \\
GHG & Greenhouse gas \\
GMST & Global mean surface temperature \\
GVA & Gross value added \\
IAM & Integrated assessment model \\
IIASA & International Institute for Applied Systems Analysis \\ 
IO-LCA & Input-output life cycle assessment \\
IPCC & Intergovernmental Panel on Climate Change  \\
JRC & Joint Research Center \\
LCA & Life cycle assessment \\
LCI & Life cycle inventory \\
LCIA & Life cycle impact assessment \\
PBA & Production-based accounting \\
PB & Planetary boundary \\
RTE & Transmission system operator \\
SSP & Shared socioeconomic pathway \\

\toprule[2pt]

\end{tabular}
}
} 
    
\end{table}

Therefore, this paper builds on and extends the original UNCASE framework to propose the UNCASE\textbf{xt} framework (read as ``UNCASE e\textbf{xt}ended''). This work brings three main contributions,
while further aligning with the AESA guidance from the European Commission's Joint Research Centre~(JRC)~\citep{bjornguidanceJRC}. 
First, we broaden the range of uncertainty sources considered in AESA and support the attribution of result variability to individual uncertainty sources. This enables AESA practitioners not only to propagate uncertainty, but also to identify which sources most influence the results. Second, we address scope consistency between estimated environmental burdens and allocated carrying capacities. To this end, we propose a classification of functional units, system boundaries, allocation paths, and corresponding enacting metrics, including new metrics introduced in this work. This helps AESA practitioners position their studies, reduce scope mismatches, and ensure consistency with the chosen functional unit. Third, we provide an open-source Python package named \texttt{pyaesa} alongside the \textit{UNCASExt} framework, enabling reproducible and systematic implementation as well as the allocation of carrying capacities from the global level to countries and sectors for all functional units introduced in this work. In addition, the framework incorporates temporal dynamics for both retrospective and prospective assessments through two distinct components of AESA: 
(i) yearly allocated shares of carrying capacities~(aSoCCs) can be calculated, with shares that may remain constant or vary over time depending on the allocation method considered; and (ii) carrying capacities themselves can be represented either as static steady-state thresholds or, for climate change, as dynamic greenhouse gas~(GHG) budgets derived from the Intergovernmental Panel on Climate Change Sixth Assessment Report~(IPCC AR6) under Shared Socioeconomic Pathway~(SSP) transition scenarios~\citep{ar6_ghg_paths, ar6}.
To illustrate the practical implementation of the proposed framework, a case study is conducted on the absolute environmental sustainability of electricity consumption in France over the period 2019--2060. The case study underlines the importance of scope matching between estimated environmental burdens and allocated carrying capacities, demonstrates the operationalization of the \textit{UNCASExt} framework and its Python implementation \texttt{pyaesa} in retrospective and prospective settings, and shows how the framework can support decision-making by assessing multiple prospective scenarios for the future French electricity mix.

\par 

The rest of this paper is structured as follows. Section~\ref{sec:literature-review} provides a literature review of uncertainty sources in AESA and a description of scope mismatch, which we identify as a fundamental limitation overlooked in the current literature. Section~\ref{sec:methods} then describes the proposed \textit{UNCASExt} framework and focuses on key methodological aspects. To illustrate the practical implementation of the proposed framework, two main results are provided in Section~\ref{sec:results}. First, the effect of functional unit mismatch on allocated shares of carrying capacities is quantitatively evaluated for the first time, demonstrating the importance of a consistent functional unit definition in environmental burden estimation and carrying capacity allocation. Then, a case study assesses the absolute environmental sustainability of electricity consumption in France from 2019 to 2060. Section~\ref{sec:discussion} then discusses limitations and perspectives of this work to finally provide concluding remarks in Section~\ref{sec:conclusion}. An overview of the comprehensive supplementary material is provided in Section~\ref{sec:appendixes} to further support transparency and reproducibility.

\section{Literature review} \label{sec:literature-review}

This section provides an overview of the gray and scientific literature on AESA. It first summarizes current harmonization initiatives, then provides a comprehensive list and classification of uncertainty sources, and finally highlights how previous studies may involve structural inconsistencies between the system boundaries and functional units used for the Absolute Sustainability Ratio~(ASR) numerator (i.e., the impact assessment results) and the denominator (i.e., the allocated carrying capacities).

\subsection{Toward harmonization of AESA studies}

AESA is structured in three main phases~\citep{bjornguidanceJRC}.
Phase~A focuses on estimating the environmental burdens of the studied activity, which relies on existing LCA standards (e.g., ISO 14040–44), or on alternative environmental accounting approaches, such as corporate environmental accounting (e.g., GHG protocol) and environmentally extended multi-regional input-output (EE-MRIO) analysis. Phase~B allocates carrying capacities by distributing global or regional environmental carrying capacities to the studied activity based on explicit allocation methods. These allocation methods are defined as combinations of sharing principles (i.e.,~distributive justice theories~\citep{ryberg2020} including acquired rights, egalitarian, utilitarian, and prioritarian), and enacting metrics \citep{bai2024}, which operationalize these principles in quantitative terms \citep{puig2025quantifying}. A wide range of such methods has been documented in the literature \citep{verhaeghe2024carrying, bjornguidanceJRC}. Finally, Phase~C interprets the results by comparing the outcomes of phases~A~and~B. This comparison is expressed through the most commonly used metric in AESA, i.e., the Absolute Sustainability Ratio (ASR) calculated for each environmental category \textit{e} as:

\begin{equation}
ASR_{e}
=
\frac{
IS_{e}
}{
\underbrace{aSoCC_{e} \times CC_{e}}_{aCC_{e}}
}
\end{equation}

where $IS_{e}$ denotes the estimated environmental burden or impact score (i.e., result of Phase~A) for environmental impact category $e$, \( aSoCC_{e} \) the allocated share of carrying capacity assigned to the studied activity, \( CC_{e} \) the corresponding global carrying capacity, and \( aCC_{e} \) the allocated carrying capacity (i.e., result of Phase~B). \( ASR_{e} \leq 1, \forall e \) indicates that the activity can be considered absolute environmentally sustainable, as it operates within its allocated carrying capacities~\citep{bjorn2020}. \par 

Nevertheless, as AESA applications have proliferated across scales and contexts, the associated terminology, methodological choices, and underlying assumptions have diversified accordingly. This increasing heterogeneity has rendered AESA outcomes difficult to interpret and to compare across studies \citep{puig2025quantifying}. In response, the recent JRC guidance~\citep{bjornguidanceJRC} seeks to provide structured and harmonized guidelines for conducting AESA studies across its three core phases~A, B and C.
In parallel, several initiatives have emerged to further harmonize AESA methods at different scales. For instance, \citep{paulillo2026} formulates recommendations on allocation methods for country-level AESA studies, while the UNCASE framework \citep{puig2025quantifying} introduces the first systematic approach for applying AESA at the country and sectoral levels under uncertainty.

\subsection{Uncertainty sources in AESA}

\begin{table*}[ht!]
\centering
\caption{Sources of uncertainty and variability in absolute environmental sustainability assessment (AESA). Although not exhaustive, the table summarizes most of the known uncertainty sources as described in the literature.} 

\label{tab:uncertainty-sources}

\begin{threeparttable}

\renewcommand{\arraystretch}{1.3}

\resizebox{1\textwidth}{!}{%
\begin{tabular}{p{0.13\textwidth} p{0.54\textwidth} p{0.15\textwidth} p{0.34\textwidth} p{0.1\textwidth}}
\toprule[2pt]
\textbf{AESA phase\newline \citep{bjornguidanceJRC}} &
\textbf{Uncertainty source} &
\textbf{Uncertainty type} &
\textbf{References} &
\textbf{Covered by\newline UNCASE\newline(original framework)} \\
\toprule[2pt]

\rowcolor[HTML]{efefef}
\multicolumn{5}{l}{\textbf{Phase~A: Estimate the environmental burdens of the studied activity (LCA focus)}} \\

Goal and scope of the LCA &
Choice of system boundaries &
Model &
\citep{hauschild2018, huijbregts1998} &
\No \\
\quad &
Choice of functional unit &
Scenario &
\citep{hauschild2018, huijbregts1998}  &
\No \\

Inventory analysis &
Inaccurate, non-representative or incomplete LCI data &
(i) Parameter \newline (ii) Spatial variability &
\citep{hauschild2018, huijbregts1998}  &
\Yes\tnote{a} \\
\quad &
Differences in yearly LCI data under different prospective scenarios (e.g., IAM-SSP-climate pathways scenarios) &
(i) Temporal variability \newline (ii) Scenario &
\citep{debortoli2025, sacchi2022}  &
\No\\
\quad &
Choice of multi-functionality resolution approach &
Model &
\citep{hauschild2018, huijbregts1998}  &
\No \\

Impact assessment &
Choice of LCIA method &
Scenario &
\citep{hauschild2018, huijbregts1998}  &
\No \\
\quad &
Inconsistent location of impact categories on the impact pathway, i.e., Driver-Pressure-State-Impact-Response (DPSIR) framework &
Variability between \newline objects &
\citep{bjorn2015introducing, vea2020framework}  &
\No \\
\quad &
Inconsistent time horizon between impact categories &
Temporal variability &
\citep{hauschild2018, huijbregts1998}  &
\No \\
\quad &
Uncertainty in lifetime of substances for the definition of characterization factors &
Parameter &
\citep{hauschild2018, huijbregts1998}  &
\No \\
\quad &
Differences in yearly characterization factors under different prospective scenarios (e.g., climate pathways) &
(i) Temporal variability \newline (ii) Scenario &
\citep{clausen2025absolute, hauschild2018}  &
\No\\
\quad &
Regional differences in characterization factors &
Spatial variability &
\citep{hauschild2018, huijbregts1998}  &
\No \\

\midrule[0.2pt]

\rowcolor[HTML]{efefef}
\multicolumn{5}{l}{\textbf{Phase~B: Allocate carrying capacities}} \\

Quantification of carrying capacities (CC) &
Level of acceptable risk when defining CC threshold\tnote{b} &
Scenario &
\citep{bjorn2020, puig2025quantifying}  &
\Yes \\
\quad &
Choice of CC control variables &
Scenario &
\citep{bjorn2020, bjornguidanceJRC}  &
\No \\
\quad &
Inconsistent location of CC control variables on the impact pathway, i.e., Driver-Pressure-State-Impact-Response (DPSIR) framework &
Variability between \newline objects &
\citep{bjorn2015introducing, vea2020framework}  &
\No \\
\quad &
Inconsistent time horizon between CC control variables, including steady-state vs. dynamic (cumulative budget) approaches &
Temporal variability &
\citep{ryberg2018, ar6}  &
\No \\
\quad &
Uncertainty in CC threshold &
Parameter &
\citep{sala2020environmental, puig2025quantifying}  &
\Yes \\
\quad &
For the specific case of dynamic (cumulative budget) CCs: differences in yearly CC thresholds under different prospective scenarios (e.g., IAM-SSP-climate pathways scenarios) &
(i) Temporal variability \newline (ii) Scenario &
\citep{ar6}  &
\No\\

Allocation of shares of carrying capacities (aSoCC) &
Choice of system boundaries &
Model &
\citep{bjorn2020, bjornguidanceJRC}  &
\No \\
\quad &
Choice of functional unit &
Scenario &
\citep{bjorn2020, bjornguidanceJRC}  &
\No \\
\quad &
Normative choice of allocation methods\tnote{c} &
Scenario &
\citep{puig2025quantifying, verhaeghe2024carrying}  &
\Yes \\
\quad &
Inaccurate, non-representative or incomplete enacting metric data &
Parameter &
\citep{hauschild2018, puig2025quantifying}  &
\Yes\tnote{d} \\
\quad &
Differences in yearly enacting metric data under different prospective scenarios (e.g., IAM-SSP-climate pathways scenarios) &
(i) Temporal variability \newline (ii) Scenario &
\citep{sacchi2022, wiebe2018}  &
\No\\
\quad &
Uncertainty in lifetime of substances for the definition of characterization factors applicable to \textit{environmental} enacting metric data &
Parameter &
\citep{hauschild2018, huijbregts1998}  &
\No \\
\quad &
Differences in yearly characterization factors, applicable to \textit{environmental} enacting metric data, under different prospective scenarios (e.g., climate pathways) &
(i) Temporal variability \newline (ii) Scenario &
\citep{clausen2025absolute, hauschild2018}  &
\No\\
\quad &
Regional differences in characterization factors, applicable to \textit{environmental} enacting metric data &
Spatial variability &
\citep{hauschild2018, huijbregts1998}  &
\No \\

\midrule[0.2pt]

\rowcolor[HTML]{efefef}
\multicolumn{5}{l}{\textbf{Phase C: Interpret results}} \\

Absolute Sustainability Ratio (ASR) &
Choice of temporal aggregation of results (year-by-year assessment vs. aggregation over a time horizon) &
(i) Temporal variability \newline (ii) Scenario &
\citep{clausen2025absolute, lejeune2026pathways}  &
\No \\
\quad &
Choice of spatial aggregation of results (region-by-region assessment vs. aggregation across regions -- incl. aggregation pre or post application of country-level prioritarian methods) &
(i) Spatial variability \newline (ii) Scenario &
\citep{paulillo2026}  &
\No \\

\bottomrule[2pt]
\end{tabular}%
}

\vspace{4pt}
\begin{tablenotes}[flushleft]
\scriptsize
\begin{minipage}{\textwidth}
\item IAM = integrated assessment model, CC = carrying capacity, EE-MRIO = environmentally extended multi-regional input-output, LCA~=~life-cycle assessment, LCI = life-cycle inventory, LCIA~=~life-cycle impact assessment, SSP = Shared Socioeconomic Pathway.
\medskip

\item \textbf{Definitions of the seven main types of uncertainty sources}~\citep{bjorklund2002, hauschild2018, huijbregts1998}: \textbf{Variability} reflects inherent heterogeneity in the studied activity and is subdivided into (i) temporal variability (changes over time), (ii) spatial variability (differences across locations), and (iii) variability between objects (differences between technologies, products, or actors, etc.). \textbf{Parameter uncertainty} arises from imperfect knowledge of quantitative input values, such as inventory data or characterization factors. \textbf{Model uncertainty} results from structural assumptions and simplifications in models, such as the handling of system boundaries and multi-functionality. \textbf{Scenario uncertainty} stems from normative choices such as the selected life-cycle impact assessment (LCIA) methods. \textbf{Epistemological uncertainty} reflects lack of relevant knowledge. \textbf{Mistakes} correspond to unintended errors such as unit conversion errors. \textbf{Relevance uncertainty} concerns, for example, the accuracy or representativeness of impact categories with respect to the area of protection. The last three sources may affect all phases of an AESA study. However, they are not explicitly listed in the table, as they are not commonly subject to systematic probabilistic assessment.
\medskip

\item[a] Only covered for the foreground system.

\item[b] Carrying capacities based on the PB framework reflect a precautionary principle \citep{bjorn2020, steffen2015planetary}, while those based on EF midpoint impact categories \citep{sala2020environmental} typically reflect a ``best estimate'' approach relying on average parameter values and focusing only on scientifically well-established impact mechanisms \citep{bjorn2020, bjorn2015introducing}. As a result, it is usually more difficult for a given human activity to comply with planetary boundary-based carrying capacities \citep{bjorn2020}.

\item[c] Including the choice of historical reference year for the acquired rights (AR) sharing principle and the responsibility period for the historical responsibility (HR) sharing principle \citep{paulillo2024, paulillo2026, ryberg2020} when they are selected.

\item[d] Non-MRIO enacting metrics (Population and GDP) are treated as deterministic. For economic enacting metrics, uncertainty is limited to inter-MRIO variability and is represented by a continuous uniform distribution derived from differences across MRIO sources. For environmental enacting metrics, uncertainty is modeled using continuous uniform probability distributions based on regional coefficients of variation (CoVs) for consumption-based accounting of GHG emissions reported by~\citep{rodrigues2018}. Since these CoVs have been estimated exclusively for greenhouse gas emissions, the same coefficients are applied uniformly across all impact categories.
\end{minipage}
\end{tablenotes}

\end{threeparttable}
\end{table*}

To enable methodological harmonization, it is essential to first ensure the consistent treatment of uncertainty in AESA results, since each of the three terms of the ASR is subject to uncertainty~\citep{hjalsted2021, ryberg2021}: (i) the estimated environmental burdens of the studied activity (e.g., via LCA); (ii) carrying capacity thresholds; and (iii) allocated shares of carrying capacities. Uncertainties arising from~(i) are common to all LCA studies, whereas uncertainties arising from (ii) and (iii) are AESA-specific \citep{bjorn2019, puig2025quantifying}. Seven main types of uncertainties have been identified in the literature on LCA-specific uncertainty sources \citep{huijbregts1998, bjorklund2002, hauschild2018}: variability, parameter uncertainty, model uncertainty, scenario uncertainty, epistemological uncertainty, mistakes, and relevance uncertainty. Building on this classification and on previous attempts to identify AESA-specific uncertainties~\citep{bjorn2019, puig2025quantifying}, the main sources of uncertainty and variability in AESA studies are summarized in \autoref{tab:uncertainty-sources}.\par 

All seven uncertainty types may occur in both Phase~A (i.e., estimation of environmental burdens) and Phase~B (i.e., allocation of carrying capacities). By contrast, Phase~C (i.e., interpretation of results) is limited to scenario uncertainty, as well as temporal and spatial variability related to the choice of whether and how to aggregate results across years and regions. 

Regarding AESA-specific uncertainties in Phase~B and~C, most sources are not commonly addressed in the literature \citep{clausen2025absolute, puig2025quantifying}. Among the sources related to the definition of carrying capacities, a limited number of studies consider the level of acceptable risk used to estimate carrying capacities (i.e., scenario uncertainty), and uncertainty in the estimated threshold itself (i.e., parameter uncertainty)~\citep{ryberg2018bring, bjorn2020, sala2020environmental}. Another source concerns the time horizon over which carrying capacities are defined. This is particularly relevant for climate change, for which carrying capacities may be defined either as static steady-state annual emission thresholds, assumed to apply indefinitely, or as dynamic cumulative GHG budgets over a specified time period~\citep{bjorn2015introducing, ryberg2018, ar6}. Regarding the allocation of carrying capacities, the most widely discussed source of uncertainty in AESA concerns the normative choice of allocation methods~\citep{bjorn2020, paulillo2026}, which reflects scenario uncertainty. Other studies consider uncertainty in enacting metric data (i.e., parameter uncertainty)~\citep{puig2025quantifying}, or differences in yearly enacting metric data under different prospective scenarios~\citep{lejeune2026pathways}. Among these sources of uncertainty, previous studies have shown that the normative choice of allocation methods has the most significant influence on AESA results \citep{ryberg2018bring}.

To contribute to the harmonization of AESA studies under uncertainty, \citep{puig2025quantifying} introduced UNCASE as the first systematic framework formalizing how uncertainties should be identified, parameterized, and jointly propagated. The primary objective of UNCASE is to enable a systematic allocation of carrying capacities at the country and sectoral levels under uncertainty. The framework provides a mathematical formalization of carrying capacity allocation and introduces probability-based AESA indicators that integrate a subset of the uncertainty sources listed in~\autoref{tab:uncertainty-sources}. 
Although UNCASE represents the most systematic proposal to date for uncertainty assessment in AESA, gaps remain regarding the range of uncertainty sources covered and the lack of attribution of result variability to individual uncertainty sources. In addition to these gaps in uncertainty propagation, AESA also faces a relevance uncertainty challenge related to the interpretation of the ASR itself, since the environmental burdens estimated in Phase~A and the carrying capacities allocated in Phase~B may be defined over inconsistent scopes. The next subsection examines these scope inconsistencies between the ASR numerator and denominator.

\subsection{Scope inconsistencies between estimated environmental burdens and allocated carrying capacities} \label{sec:LR-mismatch-asr}

The robustness and interpretability of AESA studies also require consistency between Phases~A and~B. As emphasized in the JRC guidance~\citep{bjornguidanceJRC}, the estimated environmental burdens of the ASR numerator and the allocated carrying capacities of the denominator should be defined over the same scope. Yet, mismatches can arise from three main sources in practice. The first concerns impact pathway modeling, i.e., the consistency between the methods used to characterize elementary flows in Phases~A and~B. The second concerns the accounting system boundary, with the distinction between production-based accounting (PBA) and consumption-based accounting (CBA). The third concerns the demand perimeter, with the distinction between final demand and total demand, the latter including intermediate demand between sectors~\citep{oosterhoff2023}.
In all three cases, such mismatches can undermine the conceptual validity and interpretability of the ASR because the estimated environmental burdens in Phase~A and the allocated carrying capacities in Phase~B are defined over inconsistent scopes. Yet, the following sections show that this source of relevance uncertainty remains insufficiently addressed in the existing AESA literature.

\subsubsection{Impact pathway mismatch}

The first source of inconsistency concerns impact pathway modeling. The LCIA method used to characterize the elementary flows in the ASR numerator should be consistent with the carrying capacity control variables in the denominator~\citep{bjornguidanceJRC}. Moreover, as underlined by ~\citep{puig2025quantifying}, there is also a need for consistency in the denominator between the carrying capacity control variables and the LCIA-based allocation methods, such as acquired rights (i.e., grandfathering). Misalignment of the latter was discussed in UNCASE by using available characterization factors for planetary boundaries in EXIOBASE \citep{vazquez2023level}, although recent advancements allow for aligning control variables in EE-MRIOs and LCAs \citep{yangpb2026}.

\subsubsection{Accounting system boundary mismatch: PBA versus CBA}

The second type of mismatch concerns the accounting system boundary, represented in both elements of the ASR. For the numerator, in a PBA perspective, only direct contributions are considered, corresponding to Scope~1 environmental burdens. In a CBA perspective, indirect supply-chain contributions are also included, corresponding to Scope~1, 2, and 3 burdens. The same distinction applies to the ASR denominator: PBA enacting metrics capture only what is directly generated by the studied activity, whether in terms of environmental pressures or economic value, whereas CBA enacting metrics also include indirect supply-chain contributions.

This distinction is important because numerous AESA studies estimate the numerator using process-based LCA (i.e., CBA basis), while operationalizing a utilitarian sharing principle in the denominator using economic enacting metrics such as gross value added (GVA) or gross domestic product (GDP). As emphasized by \citep{chandrakumar2020}, GVA is a PBA enacting metric because it measures only the value added directly generated by the studied activity. The same applies to GDP, which corresponds to GVA plus taxes minus subsidies on products~\citep{eurostat.EuropeanSystemAccounts2013}. Therefore, when GVA or GDP is used without adjustment together with a CBA numerator, the ASR numerator and denominator refer to different accounting system boundaries.

Such inconsistencies can be found in numerous studies, including GDP-based allocation for the oil sands~\citep{SUAREZEIROA2022} and the energy sectors~\citep{weidner2022, gebara2023national}, and GVA-based allocation for the laundry~\citep{ryberg2018bring, sherwood2022}, shipping~\citep{negri2022}, hydrogen~\citep{ehrenstein2020}, electricity~\citep{algunaibet2019, puig2025quantifying}, and broader energy sectors~\citep{balanza2025}. \citep{balanza2025} explicitly note that using GVA excludes upstream supply-chain contributions, such as material extraction, and may therefore underestimate the allocated share assigned to the studied activity. However, to the best of our knowledge, the magnitude of this underestimation has not yet been quantified by previous studies.

PBA enacting metrics may also introduce a demand perimeter mismatch because they refer to the total output of a given activity. This total output includes production serving both final and intermediate demand, irrespective of whether it is consumed within or outside the country of production. This may differ from the perimeter of the estimated environmental burdens in the ASR numerator. For example, in studies such as \citep{balanza2025, puig2025quantifying},  Phase~A targets total demand in a given region, whereas the GVA-based enacting metric used for allocation in Phase~B represents the total output of the studied sector produced in that region, regardless of where it is consumed. The scope mismatch is therefore twofold: Phase~A and Phase~B differ in the accounting system boundary and the demand perimeter. The next subsection further discusses demand perimeter inconsistencies, focusing on CBA studies.

Correcting these mismatches does not necessarily require shifting from a production-anchored to a consumption-anchored allocation. An AESA study may still aim, from a distributive justice perspective, to assign shares of carrying capacities to production rather than consumption. The requirement is instead that the enacting metric used for this production-anchored allocation be expressed over the same accounting system boundary as the ASR numerator. For studies relying on GVA, value added responsibility approaches developed outside the AESA literature~\citep{lenzen2007,gopalakrishnan2021,gopalakrishnan2022} offer a possible route by tracing value added creation through supply chains using MRIOs and the Leontief inverse. Integrating such approaches into AESA would enable retaining a production-anchored normative rationale while defining the corresponding enacting metric within a CBA system boundary.

\subsubsection{Demand perimeter mismatch: final demand versus total demand}

The third type of mismatch concerns the demand perimeter represented on each side of the ASR. Here, both the numerator and denominator may follow a CBA logic, but differ in the quantity of demand covered. This occurs when the numerator represents total demand, including intermediate demand, whereas the denominator relies on enacting metrics limited to final demand. The issue is particularly relevant for AESA studies focusing partly or entirely on B2B activities, where using final demand alone in the denominator excludes part of the activity represented in the numerator and may therefore lead to under allocation, as for example in the original UNCASE case study~\citep{puig2025quantifying}. This limitation has already been identified in the AESA literature \citep{oosterhoff2023}. \citep{oosterhoff2023} proposed an MRIO-based approach to include intermediate demand in utilitarian allocation methods by attributing to upstream sectors their indirect contribution to downstream final demand. This approach has since been applied in several studies~\citep{puig2025quantifying,aliUsingMultiregionalInput2026}; however, its outputs have been compared with final demand based allocation methods, even though they represent different demand perimeters. Moreover, as in the GVA-based cases discussed above, the approach proposed by~\citep{oosterhoff2023} targets the total output of a given sector-region pair and therefore does not allow filtering by consumption region.

Total demand CBA approaches necessarily involve overlapping responsibility perimeters. As underlined by \citep{oosterhoff2023,bjorn2023, aliUsingMultiregionalInput2026}, the sum of allocated shares across stakeholders may therefore exceed 100\% of the global carrying capacity. This is not an inconsistency, but a direct consequence of the system boundary considered. In approaches such as \citep{oosterhoff2023}, allocation is first performed on a non-overlapping perimeter, such as final demand, and responsibility is then redistributed across indirectly contributing sectors. The same underlying activity can therefore contribute to the allocated shares of several stakeholders, just as the same supply-chain burdens can be counted in several LCA numerators. The ASR denominator must mirror this overlap. Once overlapping contributions are removed, allocated shares must still sum to 100\%. \par

Overall, the literature shows that AESA is moving toward greater methodological harmonization, particularly through recent guidance documents~\citep{bjornguidanceJRC} and frameworks such as UNCASE~\citep{puig2025quantifying}. However, important research gaps remain open. Only a subset of uncertainty sources has been systematically integrated into existing frameworks, and structural inconsistencies between the ASR numerator and denominator remain common, especially regarding LCIA consistency, accounting system boundaries, and demand perimeters. Such inconsistencies can undermine the validity of AESA results and lead to misleading conclusions, which may in turn be detrimental when AESA is used to support decision-making.

This work addresses these gaps by building on UNCASE~\citep{puig2025quantifying} and extending it into the proposed \textit{UNCASExt} framework. \textit{UNCASExt} first broadens the range of uncertainty sources considered and supports the attribution of result variability to individual uncertainty sources. Second, it addresses scope consistency between estimated environmental burdens and allocated carrying capacities by introducing explicit classifications of functional units together with corresponding allocation methods, and by strengthening the treatment of temporal dynamics. Third, the \textit{UNCASExt} framework is implemented in an open-source Python package, \texttt{pyaesa}, to further support the reproducibility, validity, and comparability of AESA studies. 

A fully transparent account of the scientific advances with respect to the original UNCASE framework is provided in \appendixComparison. This includes a clear overview of the key features and limitations of the original UNCASE framework, as well as all implementation refinements and new concepts introduced in \textit{UNCASExt}, along with the main motivations for these changes.

\begin{figure*}[ht!]
    \centering
    \includegraphics[width=0.86\textwidth]{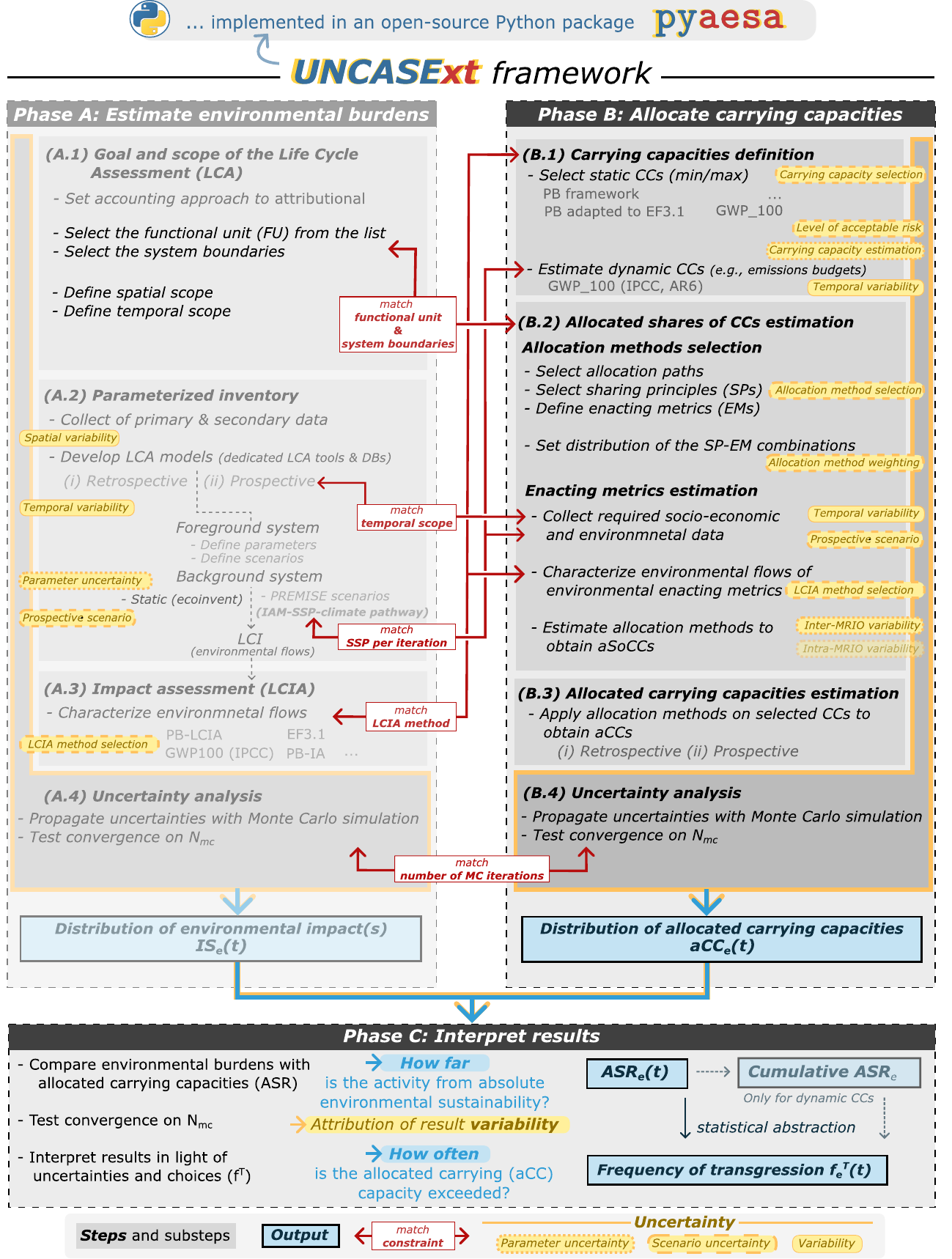}
    \caption{Overview of the \textit{UNCASExt} framework. The framework follows the three main phases of AESA, as defined in the JRC guidance~\citep{bjornguidanceJRC}.
    \textcolor{gray}{The intended workflow is as follows. First, the practitioner identifies the functional unit in Table~\ref*{tab:table-FU} that aligns with the study's goal and scope, including the targeted accounting system boundaries (PBA, CBA$_{\mathrm{FD}}$, or CBA$_{\mathrm{TD}}$). Second, the practitioner refers to \appendixEM, where allocation equations are provided for each functional unit and allocation level to compute the corresponding aSoCC. Third, the practitioner can either (i) propagate full normative uncertainty by implementing the complete set of applicable allocation methods, as in the recommended \textit{UNCASExt} workflow, or (ii) justify a constrained subset of methods, depending on the decision context and value choices. In both cases, allocation methods can either be integrated into the Monte Carlo simulation to propagate inter-method normative uncertainty or kept separate to support sensitivity analysis and explicit comparison of method-specific results. Finally, to reduce implementation risks and foster harmonization, the Python package \texttt{pyaesa} provides allocation workflows for all functional units available at $L_1$ and $L_2$, including automated processing of the required open-access data sources to calculate the corresponding allocation equations.}}  
    \label{fig:fig-method-extended-UNCASE}
\end{figure*}

\section{Methods} \label{sec:methods}

This section presents the \textit{UNCASExt} framework. First, the general workflow is described, providing specific methodological details in \appendixEM~and \appendixCC. Second, an overview of evaluated uncertainty sources is provided. Finally, a case study focusing on France's prospective scenarios for electricity consumption is defined to illustrate the \textit{UNCASExt} framework in practice.

\subsection{Overview of the proposed framework (\textit{UNCASExt})}

The \textit{UNCASExt} framework is structured in consistency with the three main phases of AESA, as illustrated in Figure~\ref{fig:fig-method-extended-UNCASE}. 
The three key contributions of the framework are: (i) broadening the set of uncertainty sources covered from Phase~A to Phase~C and attributing result variability to individual sources; (ii) ensuring scope consistency between estimated environmental burdens in Phase~A and allocated carrying capacities in Phase~B to support the interpretation of Phase~C results; and (iii) providing the open-source Python package \texttt{pyaesa}\footnote{The validation procedure for the \texttt{pyaesa} implementation is detailed in \appendixEM.} for reproducible and systematic implementation across a wide range of AESA applications. The \textit{UNCASExt} framework supports both retrospective and prospective assessments, including the definition of dynamic carrying capacities for climate change. Further methodological details are provided in the next sections.

\subsubsection{\textbf{Phase A}: Estimate environmental burdens through LCA}

\begin{table*}[t!]
\centering
\caption{Functional units covered by \textnormal{\textit{UNCASExt}} and \textnormal{\textit{pyaesa}} across allocation levels and accounting system boundaries.  Figure~\ref{fig:fig-FUs-selection}~provides a procedure to help practitioners select a functional unit.}
\label{tab:table-FU}

{\fontsize{8}{9.5}\selectfont
\renewcommand{\arraystretch}{1.2}

\begin{threeparttable}
\begin{tabularx}{\textwidth}{@{}%
  >{\raggedright\arraybackslash}p{1.2cm}%
  >{\raggedright\arraybackslash}p{11cm}%
  >{\raggedright\arraybackslash}p{2.5cm}%
  >{\RaggedRight\arraybackslash}p{0.9cm}@{}}
  
\toprule[2pt]
\textbf{Allocation level} & \textbf{Functional unit} & \textbf{Accounting system boundaries} & \textbf{Code} \\
\midrule[2pt]

$L_1$ &
Final demand of goods and services in region(s) $r_f$ in year $t$ &
$\mathrm{CBA}_{\mathrm{FD}}$ &
$L_{1.a}$
\\

$L_1$ &
Total production of goods and services by producing region(s) $r_p$ in year $t$ &
PBA &
$L_{1.b}$
\\

\midrule

$L_2$ &
Total production of goods and services by sector $s_p$ in producing region(s) $r_p$ directly supplied to final demand worldwide in year $t$ &
$\mathrm{CBA}_{\mathrm{FD}}$ &
$L_{2.a.a}$
\\

$L_2$ &
Total production of goods and services by sector $s_p$ in producing region(s) $r_p$ directly supplied to total demand worldwide in year $t$ &
$\mathrm{CBA}_{\mathrm{TD}}$ &
$L_{2.a.b}$
\\

$L_2$ &
Total production of goods and services by sector $s_p$ in producing region(s) $r_p$ in year $t$ &
PBA &
$L_{2.a.c}$
\\

$L_2$ &
Total production of goods and services by sector $s_p$ in producing region(s) $r_p$ directly supplied to final demand in region(s) $r_f$ in year $t$ &
$\mathrm{CBA}_{\mathrm{FD}}$ &
$L_{2.b.a}$
\\

$L_2$ &
Total production of goods and services by sector $s_p$ in producing region(s) $r_p$ directly supplied to total demand in  region(s) $r_c$ in year $t$ &
$\mathrm{CBA}_{\mathrm{TD}}$ &
$L_{2.b.b}$
\\

$L_2$ &
Final demand in region(s) $r_f$ in year $t$ of goods and services produced by sector $s_p$ worldwide &
$\mathrm{CBA}_{\mathrm{FD}}$ &
$L_{2.c.a}$
\\

$L_2$ &
Total demand in region(s) $r_c$ in year $t$ of goods and services produced by sector $s_p$ worldwide &
$\mathrm{CBA}_{\mathrm{TD}}$ &
$L_{2.c.b}$
\\

\midrule

$L_3$--$L_4$ &
\textit{FUs for $L_3$ to $L_4$ are provided in \appendixEM.} &
 &

\\

\bottomrule[2pt]
\end{tabularx}

\begin{tablenotes}[flushleft]
\scriptsize
\item
$L_1$~=~country or group of countries;
$L_2$~=~sector;
$L_3$~=~company;
$L_4$~=~goods or services;
FD~=~final demand;
TD~=~total demand (final $+$ intermediate);
PBA~=~production-based accounting;
$\mathrm{CBA}_{\mathrm{FD}}$~=~consumption-based accounting of final demand;
$\mathrm{CBA}_{\mathrm{TD}}$~=~consumption-based accounting of total demand.
\end{tablenotes}

\end{threeparttable}
}
\end{table*}

\begin{figure*}[ht!]
    \centering
    \includegraphics[width=1\textwidth]{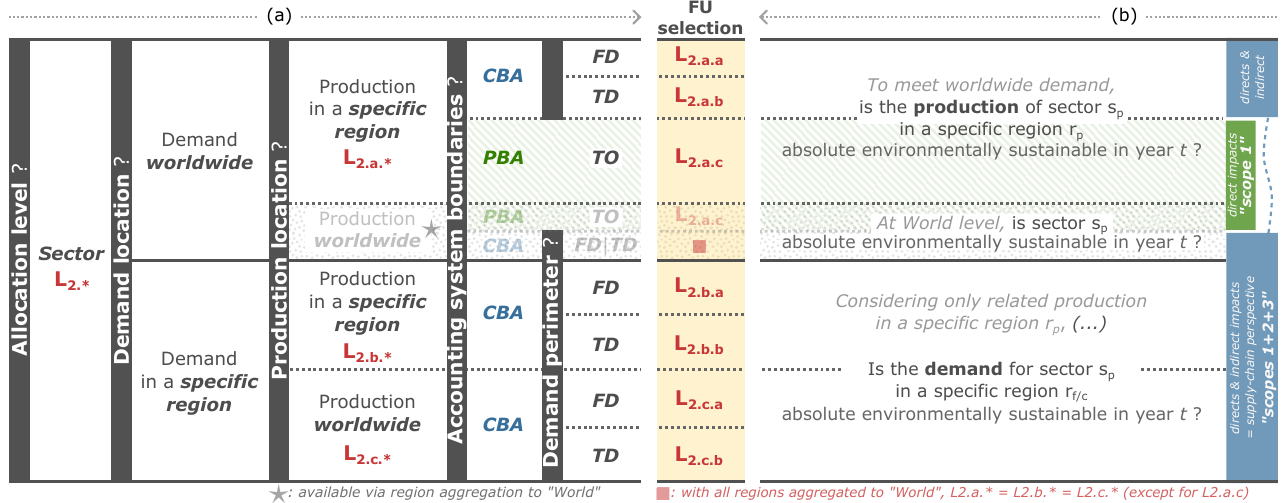}
     \caption{
     Procedure for selecting a functional unit in the \textit{UNCASExt} framework. The procedure is illustrated at the sector allocation level (L2) defined in Table~\ref{tab:table-FU}, but the same logic applies to lower allocation levels. For the sector allocation level, practitioners can identify the functional unit aligned with their study goal and scope either by (a) following the stepwise selection procedure, or by (b) directly selecting the question that most closely corresponds to their objective. In (a), practitioners first determine whether demand occurs worldwide or in a specific region, and then whether production occurs worldwide or in a specific region. They subsequently select the relevant accounting system boundaries, i.e., production-based accounting (PBA) or consumption-based accounting (CBA). CBA adopts a supply-chain perspective including direct and indirect impacts (equivalent to scopes~1+2+3), whereas PBA considers only direct impacts (equivalent to scope~1). When CBA is selected, practitioners finally determine whether the assessment covers final demand only (CBA$_{\mathrm{FD}}$) or total demand (CBA$_{\mathrm{TD}}$). For PBA, total output (TO) of the selected region is always considered by definition. Note that for the country allocation level (L1), only two functional units are available, depending on the practitioners' choice of accounting system boundaries: L1.a for CBA$_{\mathrm{FD}}$ and L1.b for PBA. }
    \label{fig:fig-FUs-selection}
\end{figure*}
 
The important novelty brought by the framework in Phase~A compared to the JRC guidance~\citep{bjornguidanceJRC} is related to the functional unit and system boundaries definition. In fact, the \textit{UNCASExt} framework emphasizes that the functional unit, temporal and spatial scopes, and system boundaries must be explicitly defined in the LCA goal and scope, since these choices impose hard scope matching constraints on Phase~B. This is illustrated in Figure~\ref{fig:fig-method-extended-UNCASE}. In particular, the demand for the activity (i.e., final demand or total demand) and the location of production and/or consumption (i.e., specific region(s) or global) should be clearly stated.
Table~\ref{tab:table-FU} reports the 21 functional units supported by the \textit{UNCASExt} framework across four levels: country or group of countries~($L_1$), sector~($L_2$), company~($L_3$), and goods or services~($L_4$). When using the \textit{UNCASExt} framework, practitioners must systematically refer to this list and select the functional unit relevant to their AESA study. To concretely help and guide practitioners in this process, Figure~\ref{fig:fig-FUs-selection} provides (a) a stepwise procedure for selecting a functional unit for L2, and (b) a reformulation of each functional unit as an AESA study goal that is easier to communicate. \par 

The rest of Phase~A remains unchanged. The goal definition of the LCA should specify the AESA study objectives and the intended audience. The environmental impacts of the studied activity can be quantified using a range of assessment methods, including the GHG protocol, input-output LCA (IO-LCA) based on EE-MRIO databases, and process-based attributional LCA~\citep{bjornguidanceJRC}. The life-cycle inventory (LCI) and its environmental flows~$F_{e}$ (i.e., elementary flows) are then classified into an impact category~$e$, and characterized through characterization factors~$CF_{F_e}$ to obtain LCIA results. The LCIA method used in Phase~A must match the categories of the control variables selected for the carrying capacities in Phase~B to ensure consistency in the AESA study.  The impact score for the environmental impact category $e$ at time $t$ is defined as $IS_{e}(t)$. The uncertainty analysis for Phase~A is performed using a Monte Carlo simulation with $N_{mc}$ iterations, whose results are then used as inputs for Phase~C.\par

\par

\subsubsection{\textbf{Phase B}: Allocate carrying capacities}

The allocated carrying capacities $aCC_e(t)$ assigned to the activity are obtained by applying the corresponding allocated shares $aSoCC_{EM,e}(t)$ to the global carrying capacities\footnote{If the carrying capacity is static (i.e., steady-state), $CC_e(t)$ is a constant~$CC_e$.} $CC_e(t)$, according to Equation~\ref{eq:general-allocation}. The following paragraphs further detail the contributions of the \textit{UNCASExt} framework with respect to these three terms.

\begin{equation}
\label{eq:general-allocation}
aCC_{e}(t) = aSoCC_{EM,e}(t) \times CC_{e}(t)
\end{equation}

\begin{figure*}[t!]
    \centering
    \includegraphics[width=1\textwidth]{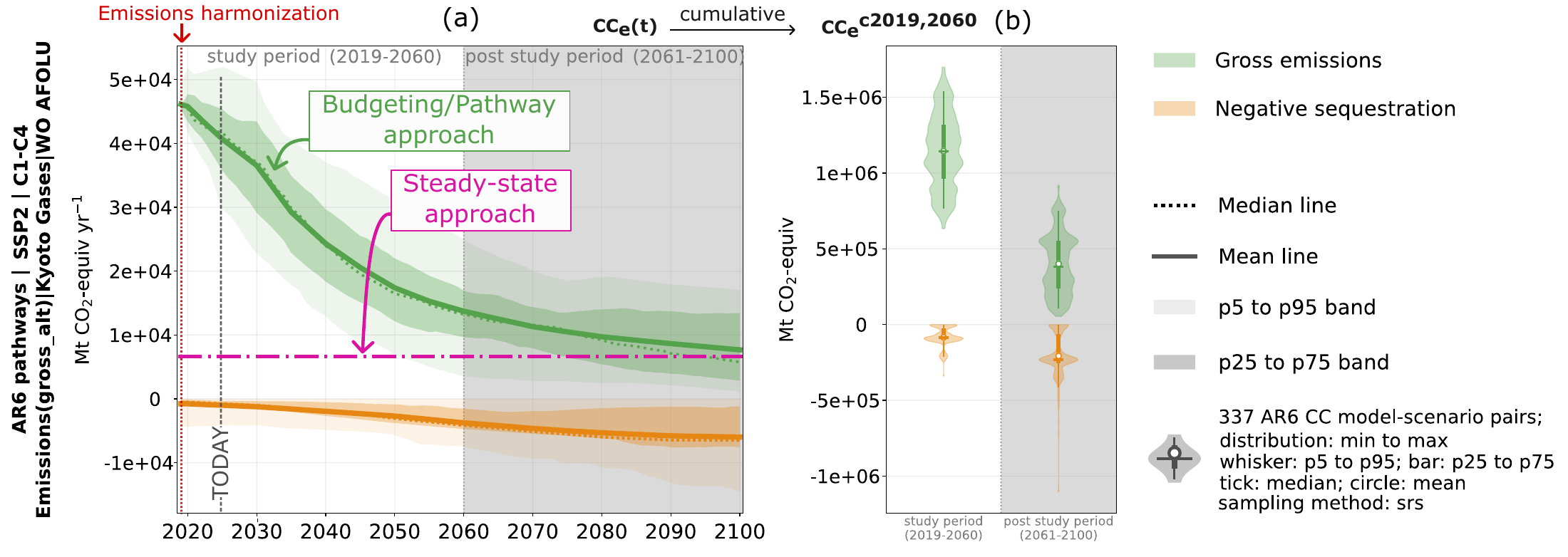}
     \caption{Alternative carrying capacity for climate change~(global) over the period 2019-2060~(illustrative example for SSP2 with GHG emissions, gross$_{alt}$, excluding AFOLU emissions). (a) Annual GHG emissions $CC_{e}(t)$ vary over time according to different IAMs and risk categories~(C1 to C4) in the budgeting approach. The static steady-state approach~(2~\textdegree C of global warming by 2100) is shown with an annual budget of 6.81~Gt\cooeq{}$\cdot$yr$^{-1}$ according to~\citep{bjorn2015introducing}, assuming an infinite time period. (b) GHG emissions budgets $CC_{e}^{c_{2019,2060}}$ based on cumulative emissions from the pathways shown in~(a). Extensive details on the definition of dynamic carrying capacities for climate change are provided in \appendixCC, including raw data collection, data processing~(e.g., AFOLU emissions and emissions harmonization), comparison with the literature, and budgets for other time periods. The workflow and figure can be reproduced using \texttt{pyaesa}.}
    \label{fig:fig-GHGbudgets-UNCASExt}
\end{figure*}

\textbf{Carrying capacities definition.} The allocation of carrying capacities for the studied activity first requires the definition of global carrying capacities. The framework therefore handles the definition of several global carrying capacities, including carrying capacities derived from the planetary boundaries (PB) framework~\citep{steffen2015planetary, richardson2023earth, kitzmann2025planetary}, and the ones adapted to the EF3.1 method~\citep{sala2020environmental, sanye2023consumption}. Details and exact values for static carrying capacities are provided in \appendixCC. An additional carrying capacity is also considered for climate change. To improve actionability for stakeholders and facilitate annual benchmarking of activities, the control variable of this alternative carrying capacity is defined in kg\cooeq{}$\cdot$yr$^{-1}$, i.e., at the pressure level in the DPSIR impact pathway framework~\citep{jabbour2014visualizing, bjorn2015introducing}. This allows to move further up the causal chain of the environmental mechanisms leading to climate change~\citep{dao2018national} to focus on \textit{pressure} (i.e., flow) rather than \textit{state} (i.e., stock) variables (i.e., GHG annual emissions rather than energy imbalance or \coo{} concentration in the atmosphere as provided in the PB framework for instance~\citep{ kitzmann2025planetary}). Both static and dynamic carrying capacities can be defined at the pressure level, as illustrated in Figure~\ref{fig:fig-GHGbudgets-UNCASExt}, allowing temporal dynamics to be included in the carrying capacity definition. Note that the dynamic carrying capacity is the global emissions budget $CC_{e}^{c_{t_0,t_f}}$ defined over the period $t_0-t_f$~\citep{gebara2023national} (i.e., the cumulative emissions) and not the annual distribution of this budget over time $CC_{e}(t)$, as captured in Equation~\ref{eq:cumulative_cc}.

\begin{equation}
\label{eq:cumulative_cc}
    CC_{e}^{c_{t_0,t_f}} = \sum_{t_0}^{t_f} CC_{e}(t)
\end{equation}

In practice, for dynamic climate change carrying capacities, GHG and \coo{} emission budgets are derived from emission pathways extracted from the International Institute for Applied Systems Analysis~(IIASA) AR6 database~\citep{byers_edward_2022_5886911}. This makes it possible to explicitly relate the definition of carrying capacities to IAMs, SSPs, and risk categories\footnote{\textit{C1}: limit warming to 1.5\textdegree C~($>$50\%) with no or limited overshoot; \textit{C2}: return warming to 1.5\textdegree C~($>$50\%) after a high overshoot; \textit{C3}: limit warming to 2\textdegree C~($>$67\%); \textit{C4}: limit warming to 2\textdegree C~($>$50\%) ~\citep{ar6}.}, which refer to average surface temperature warming by 2100~\citep{clausen2025absolute, lund2025using, lejeune2026pathways}. Nevertheless, while estimating net \coo{} and GHG emission budgets over a given time period is useful, it presents a major limitation for AESA applications. Distributive justice theories used as sharing principles in allocation procedures concern the \textit{distribution of a limited resource} (here a quantity of emissions), rather than the distribution of an \textit{obligation to sequestrate emissions}, as would be implied by negative annual emissions or negative budgets. We therefore propose splitting harmonized net emissions into gross and negative emission components. This approach is conceptually similar to that proposed in~\citep{lejeune2026pathways} for the hydrogen production sub-sector within the energy sector. However, it differs in scope, as it is intended to define global carrying capacities applicable across multiple countries and sectors, rather than to a single sub-sector. This ensures that both emission pathways \textit{and} budgets remain positive and can therefore be used as global carrying capacities for AESA allocation. The associated carbon sequestration emissions and budgets must be considered jointly with gross emissions, since the \textit{obligation} to sequestrate emissions must also be allocated to maintain the corresponding net emission budgets. Yet, the allocation of sequestration obligations remains scarcely addressed in the current AESA literature and should be further discussed in future work, also considering that sequestration often corresponds to stand-alone activities outside the supply chain of the assessed activities (e.g., direct air capture or enhanced weathering). Extensive details regarding the definition of carrying capacities in \textit{UNCASExt} are provided in \appendixCC. The definition of dynamic carrying capacities is implemented in the Python package \texttt{pyaesa}, which computes global GHG and \coo{} emission budgets and pathways for any study period between 2000 and 2100, sorted by risk category and SSP. \\

\begin{figure*}[ht!]
    \centering
    \includegraphics[width=1\textwidth]{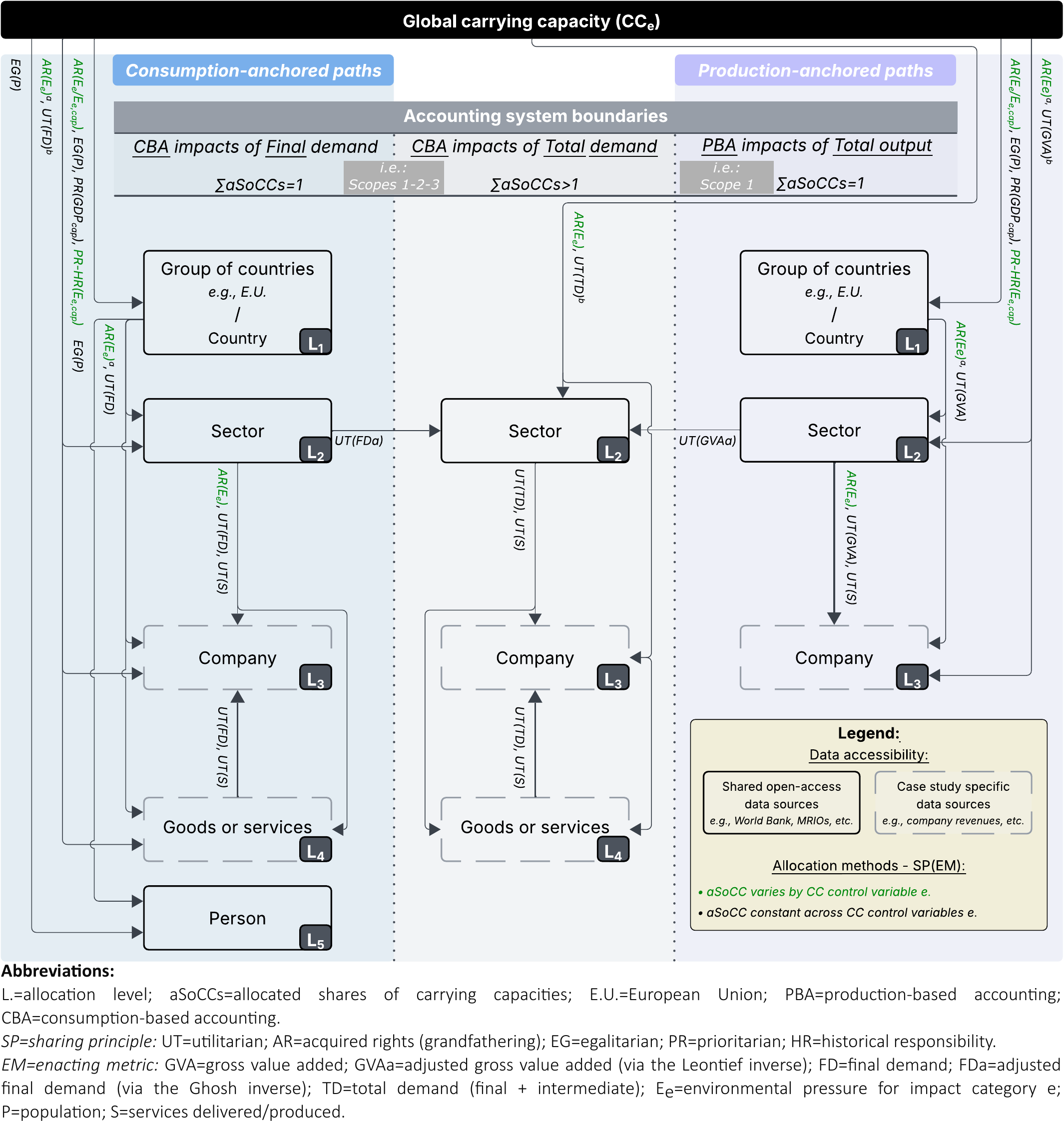}
     \caption{Overview of the accounting system boundaries, allocation paths and methods covered by the \textit{UNCASExt} framework across allocation levels. Unless stated otherwise (a), any allocation method at a given level can be combined with any allocation method at the next level along a chosen allocation path, indicated by arrows in the figure. All $L_1$ and $L_2$ allocation methods can be implemented in the Python package \texttt{pyaesa}.  Adaptation of \citep{bjornguidanceJRC}. \newline \textsuperscript{a} \textit{UT(FDa) and UT(GVAa) are adjusted variants of UT(FD) and UT(GVA) designed to reflect overlapping supply-chain propagations of utility in $CBA_{TD}$ system boundaries. Therefore, they cannot be combined within $L_2$ with AR($E_e$). Such a combination is not aligned with the intent of adjusted metrics.}}
    \label{fig:fig-asocc-paths}
\end{figure*}

\textbf{Allocated shares of carrying capacities estimation.} Then, a set of sharing principles and related enacting metrics is chosen to define allocation methods for each allocation level, down to the targeted activity. Figure~\ref{fig:fig-asocc-paths} provides an overview of the accounting system boundaries, allocation paths and methods covered by the \textit{UNCASExt} framework across allocation levels from groups of countries down to individuals. Furthermore, \appendixEM{} specifies the functional units covered by \textit{UNCASExt} for each allocation level and defines how allocation methods should be implemented for each case, including harmonized equations clarifying the treatment of final versus total demand, domestic versus foreign production, and CBA (i.e., supply-chain perspective equivalent to Scopes 1, 2, and 3 in the GHG protocol) versus PBA (i.e., production perspective equivalent to Scope 1 only in the GHG protocol) system boundaries. This formalization provides a harmonized computational framework covering 28 enacting metrics, 36 individual allocation equations, and 84 allocation paths. This ensures that the scope of the enacting metrics is consistent with the system boundaries defined in Phase~A, ultimately ensuring internally consistent AESA results. All the enacting metrics and allocation equations are implemented in \texttt{pyaesa}, including automated download and pre-processing of the required open-access data sources.

Most allocation methods formalized in \textit{UNCASExt} build on previous propositions from the AESA literature. However, scope mismatches between Phase~A and Phase~B remain widespread in current AESA studies, as discussed in Section~\ref{sec:LR-mismatch-asr}. This is particularly the case when the assessed activity requires a total demand perimeter~(B2B and mixed B2C/B2B activities) under CBA accounting system boundaries. This configuration corresponds to the central column in Figure~\ref{fig:fig-asocc-paths}.
To address this gap, \textit{UNCASExt} introduces two additional MRIO-based adjusted utilitarian enacting metrics, FDa and GVAa, which can consistently account for: (i) total demand/output rather than only final demand~(FD), and (ii) CBA accounting system boundaries rather than only PBA ones, as with gross value added~(GVA). These adjusted metrics target the same system boundaries through two allocation paths with distinct ethical anchoring. UT(FDa)-based allocation follows a consumption-anchored path, in which carrying capacities are distributed according to consumption responsibilities, whereas UT(GVAa)-based allocation follows a production-anchored path, in which carrying capacities are distributed according to production responsibilities.

At the world level, direct allocation methods such as UT(FD) and UT(GVA) allocate the full carrying capacity once across all activities: their allocated shares therefore sum to one. By contrast, UT(FDa) and UT(GVAa) start from these direct shares and propagate them through MRIO supply-chain dependencies to represent total demand/output under CBA accounting system boundaries. This propagation creates overlapping allocated shares across sectors, since a share initially assigned to final demand or value added can also be attributed to the upstream/downstream activities that enable it. Consequently, the sum of allocated shares across all activities exceeds one~\citep{oosterhoff2023,bjorn2023}. This is expected and necessary when both the ASR numerator (i.e., Phase~A) and denominator (i.e., Phase~B) are formulated on a $CBA_{TD}$ basis, since overlapping boundaries should be represented symmetrically. In Phase~A, for instance, an LCA of a studied activity may include upstream burdens that also fall within the system boundaries of other activities through Scope~2 or Scope~3 relationships.

On the consumption-anchored side, UT(FDa) extends FD as a proxy for utility by attributing to an upstream supplier the share of utility embodied in the final consumption of downstream sectors' outputs, using an upstream propagation mechanism based on the Ghosh inverse~\citep{lenzen2010}. It is conceptually close to the total demand approach proposed by \citep{oosterhoff2023}, but provides greater flexibility by enabling filtering according to the country where the studied activity's outputs are first sold, either to final demand or to a downstream sector.

On the production-anchored side, UT(GVAa) extends GVA as a proxy for utility by attributing the upstream value added required to produce the activity's total output to the activity itself. This addresses the limitation of UT(GVA), which only reflects direct value added generated by the studied activity~(i.e., Scope~1 utility), and may therefore lead to underallocation when the ASR numerator~(i.e., Phase~A) includes upstream burdens~\citep{balanza2025}. UT(GVAa) is consistent with value added responsibility approaches that propagate value added through supply chains according to upstream production dependencies~\citep{lenzen2007,gopalakrishnan2021,gopalakrishnan2022}. All details regarding the definition and computation of UT(FDa) and UT(GVAa) are provided in \appendixEM.\par 

Given that the normative choice of allocation method combinations is known to have a strong influence on AESA results \citep{ryberg2020}, all allocation methods are systematically reported together with their equations in \appendixEM. The list and classification of allocation methods are built upon the UNCASE framework \citep{puig2025quantifying}, while their formal mathematical definition extends the work of \citep{paulillo2026} from the country level down to the sector level. \par 

The main data sources used in the \textit{UNCASExt} framework for Phase~B (i.e., allocation of carrying capacities) are MRIO databases (i.e., EXIOBASE~v3.10.2~\citep{stadler2018, exiobase3102} and OECD-ICIO~v2025~\citep{oecd2023}), World Bank data for gross domestic product (GDP) and population data~\citep{worldbank}, and SSP data from the IIASA database~\citep{SEar6IIASA} in the case of prospective analysis. These data sources are selected because they provide broad temporal and spatial coverage, and have been externally validated, thereby supporting the reliability of the analysis. Relying on MRIO databases is a modeling choice primarily motivated by their ability to ensure macroscopic coherence, and therefore additivity in the computation of allocated shares, by capturing the macroeconomic structure to estimate economic and environmental data over time and across different countries. For economic enacting metrics, inter-MRIO variability is accounted for in the framework, whereas intra-MRIO variability associated with algorithms used to balance input-output tables is not available in currently available MRIO databases, hence preventing the inclusion of this uncertainty source. For environmental enacting metrics, regional coefficients of variation~(CoVs) of consumption-based carbon accounts reported by~\citep{rodrigues2018} are used to define continuous uniform probability distributions, as in the original UNCASE framework~\citep{puig2025quantifying}. Due to current data availability, these CoVs are applied to all impact categories when using EXIOBASE, which represents a limitation of the uncertainty modeling~. \appendixEM{} provides further details on data sources and uncertainty assessment for computing the \textit{UNCASExt} framework enacting metrics. \par 

When using EE-MRIOs to compute environmental enacting metrics in Phase~B, the characterization of environmental stressors (i.e., environmental flows) must be consistent with the LCIA method used for the LCA in Phase~A. For carrying capacities defined based on the PB framework, the LCIA method can therefore be the PB-LCIA method\footnote{The PB-LCIA method operationalizes the planetary boundaries within a LCA framework by defining characterization factors for about 85 elementary flows. However, the novel entities boundary is not included because the control variables are considered too immature~\citep{lund2025using}. The PB-LCIA method for Brightway2~\citep{galan2021sustainability, puig2025quantifying} has been adapted to ecoinvent~v3.10 and is available at \href{https://github.com/gpuigsamper/PB-LCIA/tree/ei_310}{gpuigsamper/PB-LCIA at ei\_310}.}\citep{ryberg2018bring} in accordance with the planetary boundaries control variables. This work relies on the characterization matrix from \citep{Yang2025, yangpb2026} for the PB-LCIA characterization of EXIOBASE~v3.10.2 environmental stressors, which has been slightly modified according to~\citep{galan2021sustainability, vazquez2023level} to have concordance with the biodiversity intactness index (BII) control variable for biosphere integrity (functional diversity). The EF3.1 method is used for carrying capacities specifically adapted to it, whereas the GWP100~(IPCC) method is used for this alternative climate change carrying capacity. Additional details are provided in \appendixCC. \par

In its current implementation of \textit{UNCASExt}, \texttt{pyaesa} allocates down to $L_2$ (sector-level). Further allocation levels ($L_3$ and $L_4$) can then be implemented if needed, as discussed in Section~\ref{sec:discussion}. \textit{UNCASExt} allows allocated shares of carrying capacities to be computed systematically over time. Population and GDP enacting metrics rely on World Bank data through 2024~\citep{worldbank} and prospective trajectories on SSP scenarios from 2025 onward~\citep{iiasa_ssp}. By contrast, MRIO-based economic enacting metrics are limited by MRIO data availability: EXIOBASE~3.10.2 covers 1995--2024 (2023 and 2024 are nowcasted)~\citep{stadler2018,exiobase3102}, while OECD-ICIO~v2025 covers 1995--2022~\citep{oecd2023}. Although prospective MRIOs have been proposed in the literature~\citep{wiebe2018, beaufils2022,cap2025}, projecting full MRIO tables is outside the scope of this work because it would require detailed supply and use tables that are not publicly available, as well as scenario-consistent alignment with the prospective \texttt{premise}~\citep{sacchi2022} background LCIs used in Phase~A. Therefore, currently \textit{UNCASExt} projects only the MRIO-based economic enacting metrics required by the allocation equations, using two strategies: regression projection and historical reuse. The regression strategy, available for non-adjusted economic enacting metrics, is inspired by the GDP scaling approach of~\citep{wiebe2018}: ordinary least squares~(OLS) level regressions\citep{greene2003econometric,seber2003linear, wooldridge2009introductory} project regional monetary totals from SSP GDP trajectories, while OLS log ratio regressions project sector and producer-region shares, preserving additivity by construction~\citep{pawlowsky2015modelling}. The method is therefore partially SSP-dependent, since SSP trajectories affect regional totals, but not relative sector, technology, or trade structures. By contrast, historical reuse is available for all economic enacting metrics and is required for UT(FDa) and UT(GVAa), because these adjusted methods depend on the full inter-country and inter-sector MRIO structure via Leontief and Ghosh inverses. Further details on data sources and projection methods are provided in \appendixEM. \\

\textbf{Allocated carrying capacities estimation.} The resulting allocated shares of carrying capacities are then applied to the global carrying capacities to eventually obtain the allocated carrying capacities for the activity, according to Equation~\ref{eq:general-allocation}. \\

\textbf{Uncertainty analysis.} Finally, Phase~B uncertainty is propagated through a Monte Carlo simulation with $N_{mc}$ iterations, following a similar approach to Phase~A. As in the original UNCASE framework \citep{puig2025quantifying}, the choice of allocation method is represented as scenario uncertainty using discrete probability distributions\footnote{Implementation in \texttt{pyaesa} allows users to either account for inter-method uncertainty within the Monte Carlo simulation or report independent results for each allocation method.}, with equal weights assigned by default to each sharing principle and to each enacting metric within a given sharing principle. \textit{UNCASExt} extends this approach by introducing a weights tree~(see \appendixEM), which makes allocation assumptions explicit and allows users to adjust the relative importance assigned to each allocation method. The resulting Phase~B outputs are then used as inputs for Phase~C.

\subsubsection{\textbf{Phase C}: Results interpretation}

The outputs of the framework are twofold and are taken from the initial UNCASE framework, i.e.,~(i)~the absolute sustainability ratio $ASR_{e,k}(t)$, and (ii)~the frequency of transgression denoted $f_{e}^{T}(t)$ as defined in Equations~\ref{eq:ASR}~and~\ref{eq:fT}.

\begin{equation}
\label{eq:ASR}
    ASR_{e,k}(t) = \frac{IS_{e,k}(t)}{aCC_{e,k}(t)} 
\end{equation}

All AESA studies rely on the ASR, which compares the environmental impacts of a studied human activity with its allocated carrying capacity per environmental impact category, $e$. In the \textit{UNCASExt} framework, it is evaluated for a single year, $t$, and for each Monte Carlo iteration, $k$, which then allows to obtain the ASR distribution. 

The frequency of transgression, as proposed in UNCASE\footnote{The initial UNCASE framework introduced the frequency of \textbf{no-}transgression denoted $f_{e}^{NT}(t)$. However, this can introduce confusion for the practitioner during the interpretation as the ASR should be minimized while the $f_{e}^{NT}(t)$ should be maximized. To ease results interpretation, we instead use the frequency of transgression $f_{e}^{T}(t)$. This is a simple statistical abstraction metric for the ASR, and both should be minimized in AESA.}~\citep{puig2025quantifying}, provides additional information as it integrates the different sources of uncertainty across all Monte Carlo iterations to ultimately determine the \textit{probability} of the activity being absolute environmentally unsustainable. In other words, the percentage of Monte Carlo runs where the ASR is greater than 1, as provided in Equation~\ref{eq:fT}. This is an essential measure of robustness that can only be provided when accounting for uncertainties~\citep{rosenbaum2017uncertainty}. It can then support result interpretation and inform recommendations in decision-making contexts. This highlights the challenge for an activity to be classified as absolute sustainable from a strong sustainability perspective, as "it must respect all the environmental issues that are relevant for the environmental sustainability objective of the assessment"~\citep{bjorn2019}. Concretely, this translates for instance into $f_{e}^{T}(t)~<~0.05$ for all planetary boundaries when protecting Holocene-like conditions~\citep{puig2025quantifying}. \par

\begin{equation}
\label{eq:fT}
    f^{T}_{e}(t) = \frac{\#^{N_{mc}}_{k=1} \big(1 - ASR_{e,k}(t) < 0\big)}{N_{mc}}
\end{equation}

The ASR and frequency of transgression metrics are mostly relevant for static steady-state analyses, where the functional unit of the assessed activity is assumed to be carried out indefinitely. In fact, the ASR evaluates the impact on the carrying capacity if the activity continues as measured in the studied year forever~\citep{guinee2022life}. This is the case when considering carrying capacities from the PB framework and the PB-LCIA method, as the LCI information on resource use and emissions to the environment is provided as mass per year, rather than mass, as in conventional LCIA methods~\citep{lund2025using}. However, extending the temporal scope and considering dynamic carrying capacities for climate change (i.e., emissions budgets) raises concerns about the relevance of an annual ASR as defined in Equation~\ref{eq:ASR}, since transitioning pathways evolve over time. In fact, if the carrying capacity is defined as a budget over a time period, an ASR computed on an annual basis may change significantly from one year to another depending on how environmental burdens evolve (e.g., infrastructure deployment) in a prospective dynamic LCA. Therefore, in the case of dynamic carrying capacities for climate change, we argue that a cumulative ASR ($ASR^{c_{t_0,t_f}}_{e,k}$) should be used, to evaluate the ratio between the cumulative environmental burdens and the cumulative carrying capacity from $t_0$ to $t_f$ for the simulation run $k$, as shown in Equation~\ref{eq:asr-cumulative}. This allows for preserving annual emissions to ease benchmarking and operationalization, while making explicit that the carrying capacity is the global emissions budget $CC_{e,k}^{c_{t_0,t_f}}$ defined from $t_0$ to $t_f$~\citep{gebara2023national} (i.e., the cumulative emissions).

\begin{equation*} 
    ASR^{c_{t_0,t_f}}_{e,k} = \frac{IS_{e,k}^{c_{t_0,t_f}}}{aCC_{e,k}^{c_{t_0,t_f}}} = \frac{\sum_{t_0}^{t_f} IS_{e,k}(t)}{\sum_{t_0}^{t_f} aCC_{e,k}(t)}
\end{equation*}

\begin{equation} 
\label{eq:asr-cumulative}
    = \frac{\sum_{t_0}^{t_f} IS_{e,k}(t)}{CC_{e,k}^{c_{t_0,t_f}} \sum_{t_0}^{t_f}  \big( \alpha_{e,k}(t) \cdot aSoCC_{EM,e,k}(t)  \big)} 
\end{equation}

\begin{equation}
\text{with}
    \begin{cases}
        CC_{e,k}(t) &= CC_{e,k}^{c_{t_0,t_f}} \cdot \alpha_{e,k}(t) \quad \forall t \in [t_0,t_f] \\
        \sum_{t_0}^{t_f} \alpha_{e,k}(t) &= 1 
    \end{cases}       
\end{equation}

where $CC_{e,k}^{c_{t_0,t_f}}$ is the cumulative budget from $t_0$ to $t_f$ for the simulation run $k$, and $\alpha_{e,k}(t)$ the annual contribution in year $t$ to the cumulative budget $CC_{e,k}^{c_{t_0,t_f}}$, i.e., the temporal shape of the pathway from $t_0$ to $t_f$.

For prospective studies using dynamic carrying capacities for climate change, the \textit{UNCASExt} framework ensures a SSP matching during the Monte Carlo simulations. This ensures that the projection of the background system used for the numerator (i.e., Phase~A) is consistent with the underlying SSP used for the enacting metrics computation in the denominator (i.e., Phase~B).

\subsection{Included uncertainty sources and propagation of uncertainties}

The systematic integration of uncertainties in AESA is one of the main motivations underlying the UNCASE and \textit{UNCASExt} frameworks. Accounting for a variety of uncertainty sources and being able to understand the impacts on AESA results and interpretation is key to robustly support decision-making and increase the credibility of AESA studies~\citep{puig2025quantifying}. Figure~\ref{fig:fig-method-extended-UNCASE} hence provides a picture of the uncertainty sources included in the \textit{UNCASExt} framework, but also sources that are not yet included or loosely captured, which is further discussed in Section~\ref{sec:discussion}. \par

Evaluating and propagating uncertainties requires the selection of an uncertainty analysis approach, typically based on analytical uncertainty propagation, Monte Carlo simulation, or pedigree matrix-based uncertainty assessment~\citep{rosenbaum2017uncertainty}. In practice, Monte Carlo simulation is selected in the \textit{UNCASExt} framework to propagate the uncertainties of both the LCA model and the allocation model. This confirms the choice initially proposed in the UNCASE framework, mostly to ease implementation as Monte Carlo is a mature probabilistic numerical simulation method that is already commonly applied in the field of LCA and is also identified as a best practice in the literature~\citep{rosenbaum2017uncertainty}. Although the accuracy of such analysis increases with the number of iterations, there is no standardized approach to define a sufficient number of iterations~\citep{rosenbaum2017uncertainty}. Convergence tests are therefore carried out to ensure that the uncertainty measures (e.g., mean or median) remain 'sufficiently stable' (arbitrary) when increasing the number of iterations~\citep{rosenbaum2017uncertainty}. 

Beyond uncertainty propagation, the contribution of individual uncertainty sources to result variability is assessed in the \textit{UNCASExt} framework through variance-based global sensitivity analysis. In particular, Sobol indices are used to decompose the output variance into contributions from individual uncertainty sources and their interactions, thereby supporting the attribution of result variability to specific sources~\citep{sobol1993,saltelli2010, jolivet2021}. 

More details regarding Monte Carlo simulations and sensitivity analysis are provided in \appendixEM{} and \appendixCC.

\subsection{Case study definition}

The \textit{UNCASExt} framework is illustrated by assessing the absolute environmental sustainability of France's electricity consumption over the period 2019-2060. To that end, we evaluated France's annual electricity consumption based on the prospective scenarios provided by France's Transmission System Operator (RTE) in \textit{Futurs Energétiques 2050} \citep{RTEscenarios}. These scenarios provide six different pathways for France's electricity production (M0, M1, M23, N1, N2, and N03), alongside three narratives for electricity demand levels (i.e., reference, sobriety, and extensive industrialization). This case study was selected to demonstrate the extended capabilities of \textit{UNCASExt} relative to the original UNCASE framework, which focused its illustration on France's electricity production in 2018\citep{puig2025quantifying}.

RTE pathways were already transformed into user-defined scenarios in \texttt{premise} to create tailor-made LCI databases\footnote{Available on GitHub at~\url{https://github.com/oie-mines-paristech/RTE_scenarios}.} \citep{RTEposter}. These user-defined scenarios systematically modify the ecoinvent 3.10.1 LCI database to align with RTE forecasts and hypotheses. Therefore, we relied on this work to model RTE scenarios until 2060. RTE scenarios for the period 2026-2060 were subsequently coupled to various IAMs and SSPs to model the evolution of the global economy, i.e., background LCIs.

The case study focused on the \textit{Reference} demand scenario from RTE, modeling two different pathways for electricity production (M0 and N03). M0 expects a high penetration of renewable energies, while N03 forecasts a strong presence of nuclear power plants. These two electricity production pathways were selected as cornerstone foreground scenarios, as they were reported to represent the extreme ends of the LCA impact scores for climate change \citep{RTEscenarios}. We acknowledge that these scenarios may not correspond to the best- and worst-case scenarios for other planetary boundaries. However, this approach was adopted to illustrate the \textit{UNCASExt} framework while limiting the number of scenarios considered for the sake of conciseness. Similar to background scenarios (IAM+SSP), the different foreground scenarios serve to evaluate the uncertainty in the foreground system by considering a range of feasible future pathways, rather than via the fine-tuned parametrization of LCIs~\citep{langkau2023, arvidsson2024}. 
Foreground scenarios were coupled to 30 background scenarios corresponding to SSP1, SSP2, and SSP5, modeled using IMAGE, REMIND, REMIND-EU, TIAM-UCL, and MESSAGE, as available in \texttt{premise} 2.3.5 \citep{mendozabeltran2020background,sacchi2022}. This diversity of IAMs, socioeconomic, and climate pathways was retained to capture variability across prospective background scenarios\citep{debortoli2025,paris2026recommendations}.

This resulted in 60 scenarios per year for the period 2026-2060. In a second step, generated scenario LCI databases were transformed into a superstructure database \citep{superstructure}.

The annual electricity consumption (i.e., total demand) in France (excluding storage) is selected as the functional unit of the LCA. This corresponds to the functional unit L2.c.b in Table~\ref{tab:table-FU}.
Electricity imports in future years were modeled using the corresponding IAM projections for the evolution of the European electricity mix. The LCIA methods were selected according to the carrying capacity definitions. Consequently, environmental impacts were estimated~(i)~based on the PB framework's control variables using the PB-LCIA method \citep{ryberg2018} via a dedicated Python package\textsuperscript{5}, and~(ii)~the IPCC 2021 characterization factors for global warming potential \citep{IPCCcfs}. Additional methodological details regarding the case study are provided in \appendixCaseStudy.

\begin{figure*}[t!]
    \centering
    \includegraphics[width=1\textwidth]{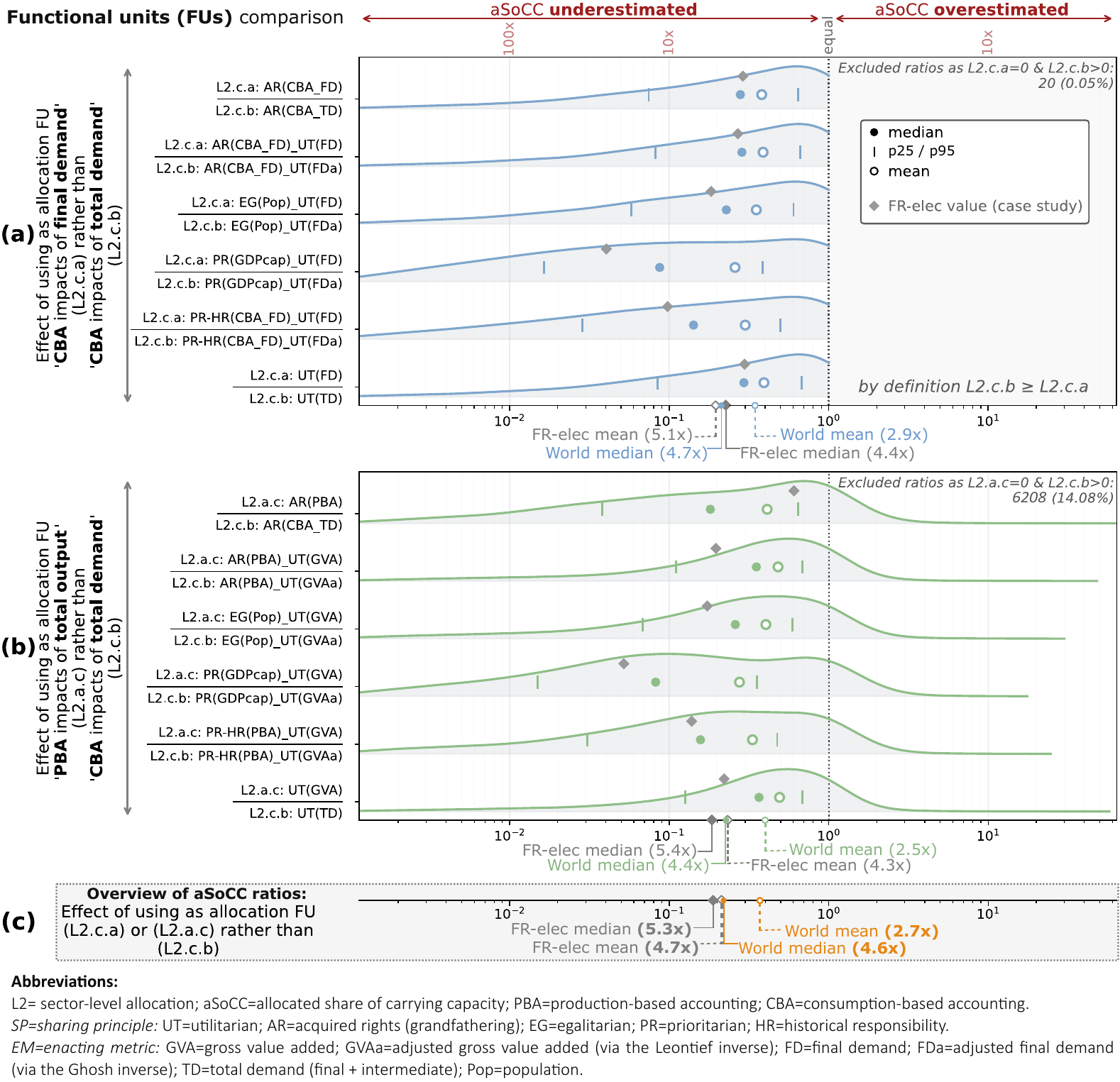}
     \caption{
    Effect of allocation functional unit (FU) choice on allocated shares (aSoCCs) across all EXIOBASE 3.10.2 ixi sector-region pairs (n = 7,350) in 2022. Ratios indicate the extent to which allocated shares are underestimated or overestimated when `PBA impacts of total output' (L2.a.c) or `CBA impacts of final demand' (L2.c.a) is used as the allocation FU rather than `CBA impacts of total demand' (L2.c.b).
    For LCIA-dependent allocation methods, the displayed impact category is GWP100. For acquired rights (AR) methods, the reference year is 1995.
    The horizontal axis is shown on a log10 scale. Ridgeline densities are displayed for values above the 5th percentile. The French electricity sector is displayed in grey.
    }
    \label{fig:fig-allocation-ratios-fus}
\end{figure*}

\section{Results} \label{sec:results}  

Two key results are provided in this section to illustrate both the \textit{UNCASExt} framework and its open-source Python implementation, \texttt{pyaesa}~(v1.2.4). First, the effect of functional unit mismatch on allocated shares of carrying capacities~(aSoCCs) is demonstrated quantitatively for the first time using all sector-region pairs available in EXIOBASE~3.10.2, thereby highlighting the importance of scope consistency between estimated environmental burdens and allocated carrying capacities. Second, the framework and \texttt{pyaesa} are applied to a full prospective AESA case study of electricity consumption in France from 2019 to 2060, with uncertainty propagation from Phase~A to Phase~C. Additional case study results are provided in \appendixCaseStudy, including a complementary retrospective AESA using IO-LCA for electricity consumption in the European Union (EU27), Brazil, Russia, India, China, and South Africa (BRICS5), and the United States, Mexico, and Canada (USMCA) over 1995--2022. These additional results further demonstrate the applicability and generalization of \textit{UNCASExt} and \texttt{pyaesa} beyond the prospective process-based LCA application.

\subsection{Effect of functional unit mismatch on allocated shares of carrying capacities}

The definition of functional units and their associated enacting metrics in the \textit{UNCASExt} framework enables a systematic comparison of allocated shares of carrying capacities~(aSoCCs) across alternative functional unit choices. In this work, we focus on two comparisons that correspond to frequent sources of mismatch identified in Section~\ref{sec:LR-mismatch-asr}: (a) mismatch between total demand and final demand perimeters, and (b) mismatch between CBA and PBA accounting system boundaries. These effects are quantified in Figure~\ref{fig:fig-allocation-ratios-fus} for six different allocation methods.\par

To generate Figure~\ref{fig:fig-allocation-ratios-fus}, \texttt{pyaesa} was applied to the complete EXIOBASE~3.10.2 ixi perimeter in 2022, comprising 7,350 sector-region pairs after aggregating all EXIOBASE electricity activities into a single electricity sector. The three functional units involved in the two mismatch comparisons ($L_{2.c.a}$, $L_{2.a.c}$, and $L_{2.c.b}$) comprise 32 compatible allocation paths in total, corresponding to 235,200 underlying aSoCC values. The 12 matched functional unit and allocation path comparisons represented in the figure yield 88,200 allocation ratio values. The distributions therefore cover the complete EXIOBASE sector-region perimeter, further illustrating the applicability of \textit{UNCASExt} and \texttt{pyaesa} across sectors and geographies.\par

Figure~\ref{fig:fig-allocation-ratios-fus}(a) shows that aSoCCs are systematically underestimated when final demand is used instead of total demand, across all allocation methods included. This effect is estimated for every sector-region pair in EXIOBASE via \texttt{pyaesa}, hence allowing for a statistical understanding of the phenomenon. The more upstream a sector is, the more pronounced the underestimation becomes, since intermediate demand accounts for a larger share of its total demand and final demand captures only a smaller share of its sector output. In extreme cases, aSoCCs can be underestimated by more than a factor 100$\times$. However, since the distributions are right-skewed, the means and medians at the world scale generally remain in the range of 1-10$\times$ underestimation. When combining all allocation methods and sector-region pairs with equal weights, this yields an underestimation of the aSoCCs by a factor of 2.9$\times$ to 4.7$\times$. \par 

Figure~\ref{fig:fig-allocation-ratios-fus}(b) provides the corresponding analysis for mismatches between PBA and CBA accounting system boundaries. The results show that aSoCCs are most often underestimated when a PBA approach is considered instead of a CBA one for the allocation of CCs. The distributions are similar to those observed for the demand perimeter mismatch, although overestimation can also occur in specific sector-region configurations. This happens for example when a sector has a high local production but exports a large share of its outputs. When combining all allocation methods and sector-region pairs with equal weights, this yields an underestimation of the aSoCCs by a factor of 2.5$\times$ to 4.4$\times$ at the world scale. \par 

Figure~\ref{fig:fig-allocation-ratios-fus}(c) summarizes the combined implications of these two mismatch types for ASR interpretation. Overall, inconsistencies between total and final demand perimeters, or between CBA and PBA accounting system boundaries, lead to aSoCC underestimation factors ranging from 2.7$\times$ to 4.6$\times$. This is a major finding because \textit{underestimating aSoCCs} mechanically \textit{overestimates the ASR}, thereby distorting the apparent gap between the environmental burden of the studied activity and its allocated share of carrying capacities. Such mistakes can affect decision-making by altering the perceived scale of transformation required for the activity to operate within its allocated carrying capacities.\medskip 

More generally, these results suggest that the common understanding in the AESA community, according to which uncertainty is mainly driven by the choice of allocation methods, should be reconsidered. As documented in Section~\ref{sec:LR-mismatch-asr}, numerous previous AESA studies have been carried out with internal mismatches between the ASR numerator and denominator. Figure~\ref{fig:fig-allocation-ratios-fus} shows that such mismatches can easily translate into variations of a factor 10$\times$ or more in ASR values. Although uncertainty due to the choice of allocation method remains significant, this work shows that ensuring consistency within the ASR is a prerequisite for robust AESA results. This is the main rationale for the systematic definition of functional units and system boundaries in the \textit{UNCASExt} framework. 

Beyond quantifying the consequences of functional unit mismatch, this analysis also demonstrates the applicability and generalization of \textit{UNCASExt} and its implementation in \texttt{pyaesa} across a wide range of sector-region combinations and allocation methods. In \appendixCaseStudy, the assessment is extended to ASRs for five $L_2$ functional units ($L_{2.a.a}$, $L_{2.a.b}$, $L_{2.a.c}$, $L_{2.c.a}$, and $L_{2.c.b}$) over the same 7,350 sector-region pairs. The five functional units comprise 54 compatible allocation paths in total, corresponding to 396,900 ASR values for GWP\_100 in 2022 using a static carrying capacity. The corresponding analyses can be fully reproduced using the code provided in the \texttt{AESA\_case\_studies} repository at~\url{https://github.com/AESAtoolkit/AESA_case_studies}.

\subsection{Case study results}

\begin{figure*}[t!]
    \centering
    \includegraphics[width=0.84\textwidth, angle=0]{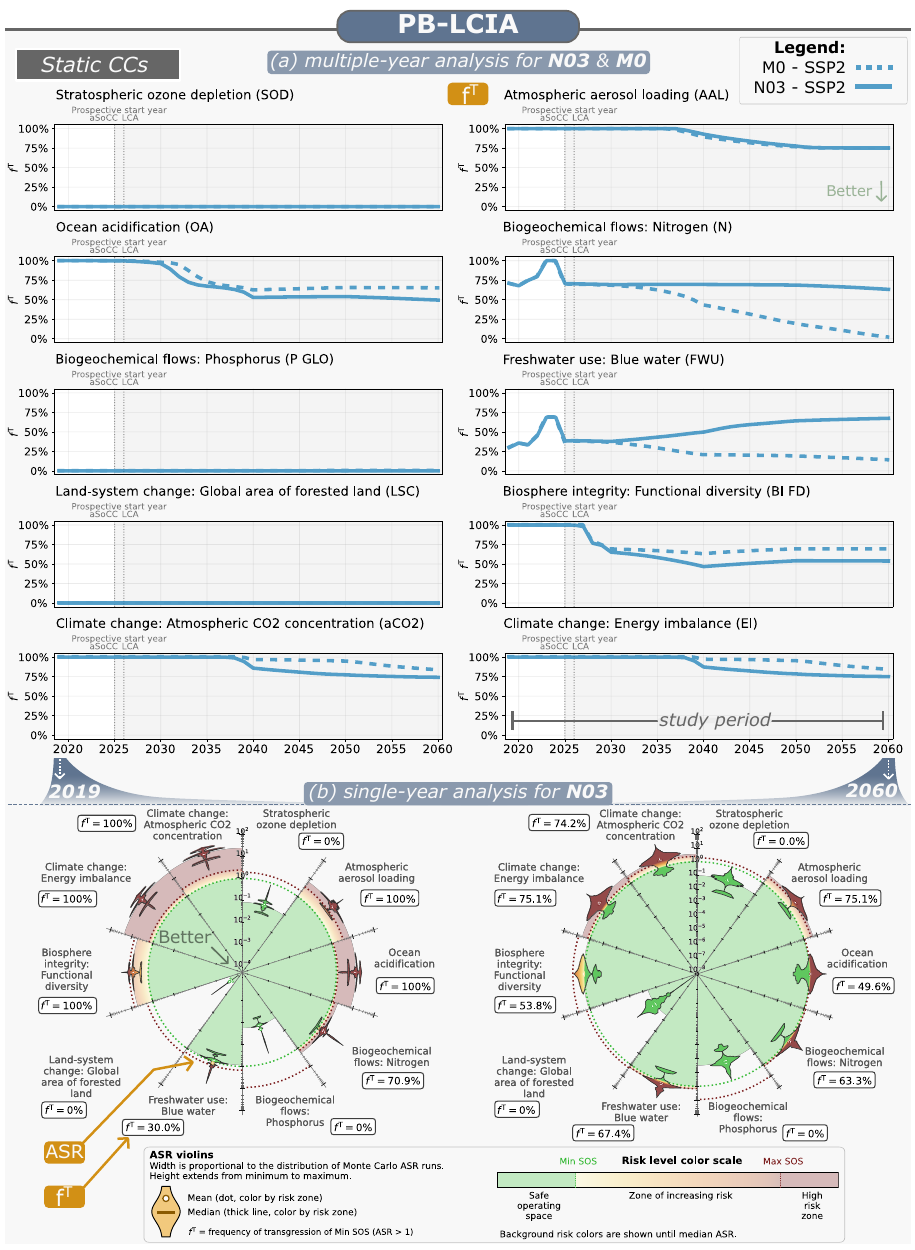}
     \caption{Case study results (with a focus on SSP2) for PB-LCIA. (a) The frequency of transgression $f^{T}$ is provided for all years in the study period (multi-year analysis) for N03 and M0, capturing the trend over time and the statistical information of uncertainty sources included via the Monte Carlo simulations. (b) Single-year analysis is also provided for the first and last year of the study period, depicting in the polar plot the statistical distribution of the ASR. \textit{Reminder}: The $f^{T}$ metric answers the question "how often is the allocated carrying capacity (aCC) exceeded across all Monte Carlo runs?" whereas the ASR metric answers the question "how far is the activity from absolute environmental sustainability?". \textit{Note:} these are only some examples of the results that can be obtained via \texttt{pyaesa}. More figures can be obtained for single- and multiple-year analysis, for both deterministic and uncertainty approaches, and for each allocation method selected. Results for other SSPs (SSP1 and SSP5) are provided in \appendixCaseStudy.
    }
    \label{fig:fig-results-pblcia}
\end{figure*}

\begin{figure*}[t!]
    \centering
    \includegraphics[width=0.76\textwidth, angle=0]{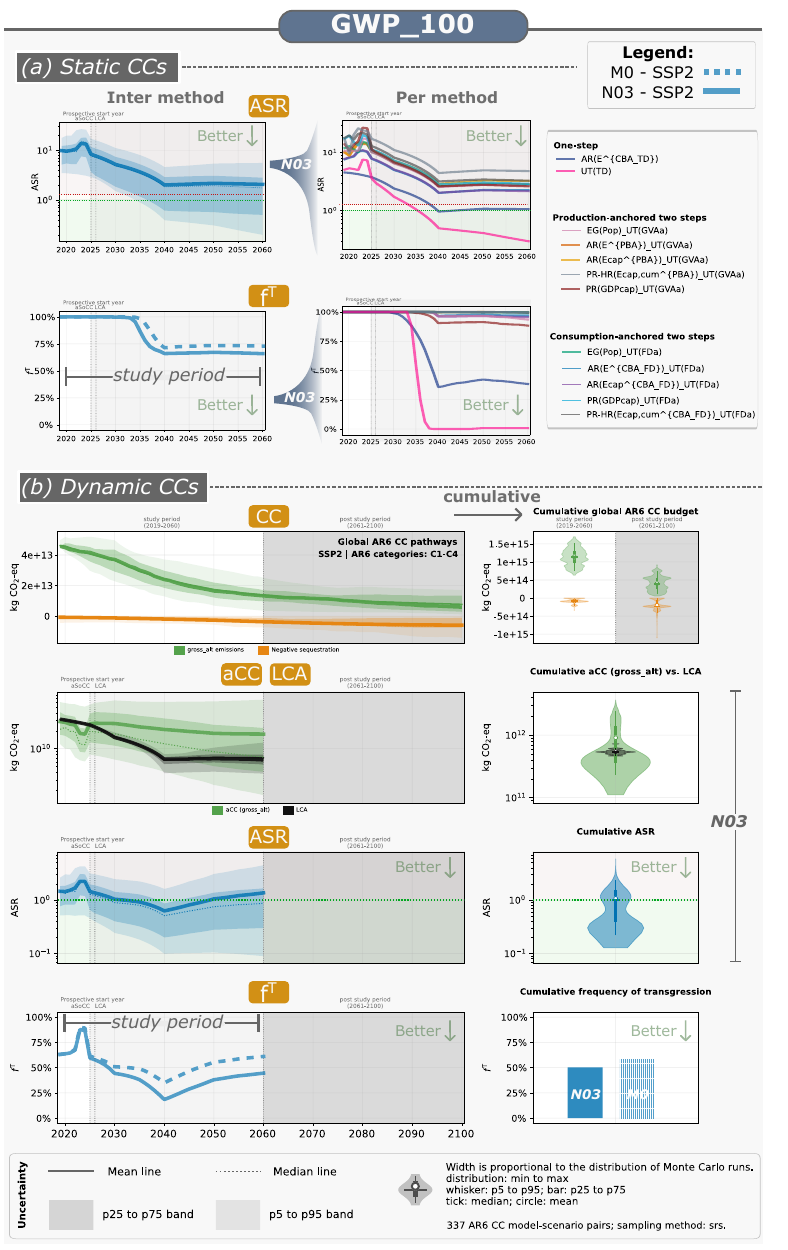}
     \caption{Case study results (with a focus on SSP2) for GWP\_100 (climate change), considering both (a) static and (b) dynamic carrying capacities. The \texttt{premise} Python package~\citep{sacchi2022} is used for generating prospective LCA results (Phase A).  The ASR and the frequency of transgression $f^{T}$ are provided for all years in the study period, capturing the trend over time and the statistical information of uncertainty sources included via the Monte Carlo simulations. \textit{Reminder}: The ASR metric answers the question "how far is the activity from absolute environmental sustainability?" whereas the $f^{T}$ metric answers the question "how often is the allocated carrying capacity (aCC) exceeded across all Monte Carlo runs?".  \textit{Note:} these are only some examples of the results that can be obtained via \texttt{pyaesa}. More figures can be obtained for single- and multiple-year analysis, for both deterministic and uncertainty approaches, and for each allocation method selected. Results for other SSPs (SSP1 and SSP5) are provided in \appendixCaseStudy.
    }
    \label{fig:fig-results-gwp100}
\end{figure*}

The results obtained with \texttt{pyaesa} when applying the \textit{UNCASExt} framework on the case study are depicted in Figure~\ref{fig:fig-results-pblcia} and Figure~\ref{fig:fig-results-gwp100} for two LCIA methods, i.e., PB-LCIA and GWP\_100. For the sake of conciseness, Figure~\ref{fig:fig-results-pblcia} and Figure~\ref{fig:fig-results-gwp100} focus on the N03 RTE scenario for the foreground system, and SSP2 as the background system scenario. The comparison with the M0 RTE scenario is presented for the frequency of transgression $f_{e}^{T}(t)$. Results for other SSPs (SSP1 and SSP5) are provided in \appendixCaseStudy. \par

Figure~\ref{fig:fig-results-pblcia} shows the evolution over time of the $f_{e}^{T}(t)$ for the N03 and M0 scenarios for France's electricity consumption. According to the results, France's electricity consumption mix only respected the planetary boundaries for land system change (LSC), phosphorus biogeochemical flows (P-GLO), and stratospheric ozone depletion (SOD) in 2019, in line with previously reported results for 2018~\citep{puig2025quantifying}. These planetary boundaries remain below the allocated carrying capacities throughout the evaluated period. The frequency of transgression $f_{e}^{T}(t)$ decreases for most planetary boundaries, indicating that the N03 scenario generally improves its absolute sustainability performance over time. Yet, all scenarios fail to achieve absolute sustainability across all categories by 2060. Environmental impacts on ocean acidification~(OA), atmospheric CO$_2$ concentration~(aCO2), energy imbalance~(EI), biosphere integrity functional diversity~(BI-FD), and atmospheric aerosol loading~(AAL) move from fully exceeding their allocated carrying capacities in 2019 to partially complying with them by 2060. For example, the $f_{e}^{T}(t)$ for BI-FD in the N03 scenario passes from 100\%  in 2019 to 54\% in 2060. However, the $f_{e}^{T}(t)$ remains above 50\% across all these planetary boundaries, suggesting that even if the sector's metrics improve, France's electricity consumption (FU: L2.c.b) will still exceed its allocated carrying capacities by 2060. The freshwater use (FWU) planetary boundary is the only category for which the $f_{e}^{T}(t)$ increases over time in the N03 scenario, indicating that the sector is clearly moving further away from its allocated share for that category.

The absolute sustainability performance of the electricity production sector is additionally evaluated using the distribution of the ASR. Figure~\ref{fig:fig-results-pblcia} presents two violin plots representing the ASR of the N03 RTE scenario for each planetary boundary in 2019 and 2060. These two snapshots confirm that, although the sector improves overall, most planetary boundaries remain transgressed in 2060~(6/10), with median ASR values ranging from 1.1 to 4.8 across the transgressed categories\footnote{This range of ASR values may seem small compared to previous AESA studies considering different allocation methods. Yet, Figure~\ref{fig:fig-allocation-ratios-fus}(c) points out that a mismatch in the ASR due to a wrong choice of allocation FU may lead to an aSoCC underestimation from a factor 4.4-5.1$\times$ for this specific case study. This would result in (erroneous) ASR values ranging from 4.4 to 24.5.}. These findings suggest that demand-side measures, such as reducing electricity demand, should be considered alongside the technological changes evaluated in this study. For both years, the findings indicate large variability of the ASR within each boundary (i.e., spanning up to three orders of magnitude).

The variance-based Sobol analysis is illustrated in Table~\ref{tab:sobol-gwp100} for GWP\_100 static carrying capacity. For this impact category, allocation method uncertainty is the dominant source of variability, with a weighted total-order Sobol index of 0.99 in 2019 and 0.69 in 2060. Its relative contribution decreases over time as prospective uncertainty sources become more influential, with background system LCA scenario uncertainty and allocation projection uncertainty reaching weighted total-order indices of 0.35 and 0.18, respectively, in 2060.

\begin{table}[t]
\centering
\caption{Variance-based analysis for 2019 and 2060 results via Sobol indices for scenario REF\_N03 and SSP2 (GWP\_100, static CC). \\}
\label{tab:sobol-gwp100}

\resizebox{\columnwidth}{!}{%
\renewcommand{\arraystretch}{0.9}

\begin{tabular}{@{}lcccc@{}} 

\bottomrule[2pt]
\textbf{Source name} & \textbf{Year} & \textbf{SSP} & \textbf{Variance} & \textbf{Variance} \\

& & \textbf{scenario} & \textbf{weighted\_S1} & \textbf{weighted\_ST} \\

\bottomrule[2pt]

acc::asocc::inter\_method\_uncertainty & 2019 & - & \textbf{0.94} & \textbf{0.99} \\
acc::asocc::inter\_mrio\_uncertainty & 2019 & - & 0.00 & 0.01 \\
acc::asocc::lcia\_uncertainty & 2019 & - & 0.01 & 0.03 \\
acc::asocc::projection\_uncertainty & 2019 & - & 0.00 & 0.00 \\
acc::asocc::reference\_year\_uncertainty & 2019 & - & 0.00 & 0.01 \\
external::lcia::rte\_reference\_n03\_ssp2 & 2019 & - & 0.00 & 0.00 \\

\hline

acc::asocc::inter\_method\_uncertainty & 2060 & SSP2 & \textbf{0.46} & \textbf{0.69} \\
acc::asocc::inter\_mrio\_uncertainty & 2060 & SSP2 & 0.00 & 0.04 \\
acc::asocc::lcia\_uncertainty & 2060 & SSP2 & 0.00 & 0.02 \\
acc::asocc::projection\_uncertainty & 2060 & SSP2 & 0.04 & 0.18 \\
acc::asocc::reference\_year\_uncertainty & 2060 & SSP2 & 0.00 & 0.01 \\
external::lcia::rte\_reference\_n03\_ssp2 & 2060 & SSP2 & \textbf{0.27} & \textbf{0.35} \\

\bottomrule[2pt]

\end{tabular}
}

\end{table}

Extending the analysis to other impact categories with PB-LCIA shows that allocation method choice remains the largest source of result variability, with total-order indices ranging from 95\% to 99\% in 2019 and from 64\% to 97\% in 2060. These results align with previous findings~\citep{ryberg2018bring,puig2025quantifying}, while providing for the first time a variance-based quantification of allocation uncertainty. This shows that its contribution is impact category dependent, since the spread between LCIA-based and non-LCIA-based allocation methods is itself impact category dependent.

Figure~\ref{fig:fig-results-gwp100} focuses on climate change (GWP\_100) and confirms the findings from Figure~\ref{fig:fig-results-pblcia}, i.e., a slight improvement for the climate change category by 2035-2040. The improvement of the $f_{e}^{T}(t)$ is limited to about 25\%-30\%, depending on the choice of static or dynamic carrying capacity, thereby leading to the same conclusion regarding the exceedance of the planetary boundary for climate change. Figure~\ref{fig:fig-results-gwp100} also shows that only considering a one-step utilitarian allocation method for the GWP\_100 static carrying capacity would lead to the conclusion that the electricity production sector is environmentally sustainable from a strong sustainability perspective before 2040. This result is similar to the findings of the UNCASE framework~\citep{puig2025quantifying}, which also identified the one-step utilitarian approach as yielding the largest share. All other allocation methods show opposite conclusions, with ASRs diverging from a factor of 4$\times$ up to a factor of more than 10$\times$ in Figure~\ref{fig:fig-results-gwp100}. 

Providing results for GWP\_100 both for static and dynamic carrying capacities allows to analyze the effect of carrying capacity selection. Although both approaches show an improvement of $f_{e}^{T}(t)$ until 2040, two key differences appear. First, the ASR and the $f_{e}^{T}(t)$ are found to be better in the dynamic approach, i.e., $f_{e}^{T}$ down to 25\% in the dynamic approach, while at best above 60\% in the static approach. This is simply explained by the fact that annual greenhouse gas emissions are significantly higher during the early decades of the 21st century with a dynamic carrying capacity as they are aligned with historical emissions, whereas the steady-state CC of 6.81 Gt\cooeq{}~$yr^{-1}$ is constant over time and leads to a smaller allocated carrying capacity (see Figure~\ref{fig:fig-GHGbudgets-UNCASExt}). Second, while the ASR and the $f_{e}^{T}(t)$ reach a plateau in the static approach, they continue to evolve in the dynamic approach. This behavior arises because the environmental burden (i.e., Phase A) stabilizes around 2040, whereas the dynamic CC continues to decrease in accordance with AR6 pathways (C1–C4), consistent with limiting global warming to below 2°C by 2100. These findings highlight (i) the effect of temporal dynamics, and (ii) the importance of considering the post-study period in the interpretation, as it shows that further efforts will be required between 2060 and 2100 for the sector to move within its allocated share of carrying capacity. Additionally, the evolution of the ASR using dynamic carrying capacities shows that absolute sustainability can be (on average) achieved in given future years, although it cannot be achieved over the entire studied period. These findings support the idea of establishing annual decarbonization targets but highlight the need to further reduce environmental impacts over time as emission budgets shrink. In this sense, the cumulative ASR and frequency of transgression indicators appear as necessary to complement time-specific AESA studies.

Further details supporting the interpretation of the ASR results are provided in \appendixCaseStudy. For Phase A, \appendixCaseStudy\ presents a decomposition of the LCA impact scores of the evaluated scenarios in 2019 and 2060 by their main contributors for each PB, clarifying the main drivers of the observed environmental impacts. For Phase B, it reports the temporal evolution of aSoCCs and aCCs, providing further insight into the changes in the ASR denominator over time. In addition, \appendixCaseStudy\ presents a complementary retrospective AESA study based on IO-LCA for electricity consumption ($L_{2.c.b}$) in the EU27, BRICS5, and USMCA over 1995--2022, with environmental impacts characterized through PB-LCIA. Mean ASRs indicate overshoot of six to seven of the ten PB control variables over the studied period. This retrospective application illustrates the applicability of \textit{UNCASExt} and \texttt{pyaesa} across temporal and spatial scopes as well as an IO-LCA-based Phase~A. Detailed ASR and $f^{T}$ trajectories are provided in \appendixCaseStudy, and the complete analysis can be reproduced using the code provided at~\url{https://github.com/AESAtoolkit/AESA_case_studies}.

\section{Discussion} \label{sec:discussion}

This paper investigates fundamental gaps in AESA and proposes important developments both in terms of methodology \textit{and} practical implementation. \par 

On the methodological aspects, the paper demonstrates the critical importance of scope consistency between estimated environmental burdens~(i.e., Phase~A) and allocated carrying capacities~(i.e., Phase~B). As shown in the results, internal mismatches within the ASR can substantially affect outcomes, making scope consistency a prerequisite for robust AESA interpretation before uncertainty sources are propagated from Phase~A to Phase~C. By formalizing scope consistency, the \textit{UNCASExt} framework helps AESA practitioners position their studies, reduce scope mismatches, and ensure consistency with the chosen functional unit.
Then, the \textit{UNCASExt} framework broadens the uncertainty sources considered from Phase~A to Phase~C and combines uncertainty propagation with variance-based analyses. This enables practitioners not only to quantify uncertainty but also to identify which uncertainty sources most influence the results via variance-based analyses. This attribution of result variability is a key step forward for result interpretation and decision-making. \par

On the practical implementation aspects, this work enables for the first time reproducible and systematic AESA implementation through \texttt{pyaesa}. The package implements the \textit{UNCASExt} framework presented in this work and demonstrates its functionalities in Section~\ref{sec:results}. The package was developed as an open-source tool to enable future improvements based on community needs, practices, and consensus.

Nevertheless, this work does not address all current AESA research gaps at once. Therefore, the rest of this section discusses the main limitations of the framework and outlines directions for future work. Perspectives are then presented for the adoption of the framework by the scientific community. Limitations specific to the case study are discussed in \appendixCaseStudy.

\subsection{Limitations of the UNCASExt framework and future work suggestions}

A first limitation of the framework is that it can be applied only to sectors that are available in (EE-)MRIO database classifications. MRIO databases enable the systematic estimation of economic and environmental enacting metrics for a wide range of sectors and countries, but they restrict the analysis to sectors and countries that are available in their classification. In practice, this can be a bottleneck because metrics estimation can only be carried out if the sector is properly captured in the MRIO classification. Some sectors are well defined (e.g., the electricity sector), while others are poorly represented or do not exist at all. Nevertheless, this limitation is well known in AESA studies, and the \textit{UNCASExt} framework helps address it through \texttt{pyaesa}, which enables users to disaggregate regions and/or sectors on a yearly basis via \texttt{pymrio}~\citep{stadler2021pymrio}. However, disaggregation requires external data to split environmental or economic flows, which complicates implementation and introduces additional uncertainties~\citep{delatorre2026, badr2026comprehensive, badr2026satellite}. In fact, the authors of \citep{badr2026comprehensive} point out in their recent comprehensive review that (dis)aggregation procedures are one of the main sources of uncertainty in MRIO modeling, together with data source selection, and assumption-based decisions in balancing and environmental extensions. In practice, (dis)aggregation is particularly challenging for pluriannual assessments, where consistent time series of disaggregation keys are not always available outside MRIO databases. When broader system boundaries are required, \texttt{pyaesa} also enables users to aggregate regions and/or sectors. \\

A second limitation is related to the treatment of final demand in MRIO databases, with two implications for the allocation of carrying capacities. (i) Gross fixed capital formation is recorded as final demand although capital goods are used in production processes. This affects final demand based economic and environmental enacting metrics at both country and sector levels. It does not create an additivity issue, since the corresponding economic and environmental flows remain accounted for. Rather, it creates an attribution issue: capital goods production is attributed to investment demand instead of being reallocated to the sectors that use these capital goods to produce goods and services over time. Capital endogenization, based on capital use matrices when available~\citep{sodersten2018, andrieu2024}, could contribute to addressing this limitation, although the temporal treatment of capital formation and use remains a methodological challenge. (ii) Direct environmental extensions associated with final consumption are not attributed to any sector. This includes, for example, fossil fuel combustion by households for private mobility or heating. This issue is specific to sector-level MRIO-based environmental enacting metrics, as these extensions are captured at the country level but remain unassigned across sectors within each country. Thus, depending on the impact category, sector-level allocated shares within a country may sum to less than 100\%, because part of the country-level environmental pressure is recorded directly for final consumption rather than for producing sectors. Redistributing these flows to the activities driving them could improve sector-level allocation, especially for air emissions and water consumption~\citep{serrano2025communicating}.\\

A third limitation concerns the temporal scope of (EE-) MRIO databases, which are only available for retrospective years. \textit{UNCASExt} currently relies on regression projection or historical reuse rather than on fully projected MRIO tables. Although prospective MRIOs have been proposed~\citep{wiebe2018, beaufils2022,cap2025}, their use in prospective AESA would require strict alignment between: (i) the projected world state represented by the inter-country and inter-sector economic flows of the MRIO tables used for allocation in the ASR denominator~(i.e., Phase~B), and (ii) the projected world state underlying the prospective life-cycle inventories used in the ASR numerator~(i.e., Phase~A), generated with \texttt{premise}~\citep{sacchi2022}. Without such alignment, environmental burdens (i.e., Phase~A) and allocated shares (i.e., Phase~B) may rely on inconsistent representations of future technologies and worldwide value chains. Additionally, for prospective assessments, changes modeled in the foreground scenarios of the studied activity, such as demand levels or production processes, may affect economy-wide trade and production structures and should therefore be reflected in the MRIO world state used to compute allocated shares through MRIO-based enacting metrics. Addressing these challenges constitutes a major research area in itself and therefore falls outside the scope of this paper. Yet, resolving them would considerably strengthen the prospective capabilities of \textit{UNCASExt} and \texttt{pyaesa}. \\

A fourth limitation concerns the concept of prospective allocation itself. In the proposed framework, prospective allocated shares of carrying capacities (aSoCCs) at  $L_2$ (sector-level) are mainly implemented through acquired rights~(AR) and utilitarian~(UT) sharing principles, optionally combined with $L_1$ country-level weights coming from egalitarian, prioritarian, or AR sharing principles. AR computes allocated shares from a selected reference year and keeps them constant over time, while UT relies on projected economic enacting metrics that reproduce historical trends. Such approaches allow prospective aSoCCs to be computed, but they may introduce path dependencies and risk reinforcing existing socio-economic structures rather than supporting the transformative changes required to move from current systems toward desirable futures that comply with carrying capacities. This echoes broader critiques of forecasting-based approaches in futures studies~\citep{robinson1990futures} and suggests that complementary approaches, especially backcasting, should be investigated to derive prospective allocated shares.

More fundamentally, most allocation methods in AESA can distribute carrying capacities according to different distributive justice theories for a given representation of the economy, but they do not define which future supply chains, consumption patterns, or production systems should be considered desirable. This remains a value-based and societal choice that cannot be substituted by allocation equations alone. Recent discussions therefore suggest that prospective AESA should move toward approaches that explicitly define desirable future states, including sufficientarian perspectives grounded in the fulfillment of basic human needs~\citep{rao2018, paulillo2024, bjorn2026}. However, current sufficientarian implementations cannot be systematically applied across countries and sectors as they require context-specific definitions of needs and future socio-technical systems~\citep{paulillo2024, kromand2025}. Moreover, they currently target only final demand consumption and therefore do not directly address upstream activities.

A promising research direction is therefore to model sufficientarian scenarios within prospective MRIO databases. This could enable the representation of desirable future supply chains across countries and sectors, covering different demand perimeters and accounting system boundaries. Such scenario-based MRIOs could then be used within the \textit{UNCASExt} framework and implemented in \texttt{pyaesa} to compute prospective allocated shares that are not only extrapolated from historical structures, but also derived from explicitly defined sets of desirable futures.\\

A fifth limitation is related to the definition of carrying capacities. The proposed framework considers the definition of \textit{global} carrying capacities, thereby assuming that the available carrying capacity is a global resource to be shared. While this assumption is valid at the Earth-system level, it can become limiting at lower spatial scales because global carrying capacities do not always capture local environmental conditions. For instance, although carrying capacities for freshwater change (blue and green) can be quantified with \textit{global} control variables as proposed in the planetary boundaries framework~\citep{richardson2023earth}, this does not reflect regional differences in water availability, scarcity, quality, or seasonal hydrological cycles, for example, between France and India. Future work could extend \textit{UNCASExt} to local carrying capacities by: (i) redefining the spatial scope of carrying capacity allocation; (ii) adapting enacting metrics to account for local carrying capacities embodied in imports and exports; (iii) evaluating environmental impacts with greater spatial granularity; and (iv) revisiting the underlying sharing principles to ensure coherence with distributive justice theories for local resources. The adjusted allocation method UT(GVAa) provides a first step in this direction by tracing the locations where value added is generated, either directly or through upstream dependencies, and linking these contributions to the environmental pressures occurring in those locations. \\

Finally, to the best of our knowledge, the \textit{UNCASExt} framework integrates the widest range of uncertainty sources across the three phases of AESA. Nevertheless, additional sources could be incorporated in future work to further increase uncertainty coverage. Suggestions are provided in \appendixEM~(Table~B.9). Examples include the use of regionalized LCI data and characterization factors, although these remain emerging in LCA tools, databases, and practice. Another example is the use of coefficients of variation~(CoVs) derived from consumption-based carbon accounts reported by~\citep{rodrigues2018} for climate change as a proxy for LCIA uncertainty across all impact categories, as done in the original UNCASE framework. However, to the best of the authors' knowledge, no alternative providing impact-category-specific CoVs has since become available in the scientific literature. Rather than overlooking this uncertainty source, it was therefore retained in \textit{UNCASExt} and implemented in \texttt{pyaesa} in a way that allows it to be easily updated once impact category specific CoVs become available. This again shows the added value of an harmonized \textit{and} flexible Python implementation for AESA studies. A third example concerns the residual mismatch between the IAMs used in \texttt{premise} to evaluate the ASR numerator in prospective LCA and those used to define GWP100-based dynamic climate change carrying capacities, as provided in Equation~\ref{eq:asr-cumulative}. Although the proposed approach directly contributes to formalizing the definition of dynamic carrying capacities for climate change, it also highlights the need for future research on the matching of temporal scopes. For instance, the integration of non-linearity for both environmental burden and carrying capacity definitions should be further investigated, as a cumulative impact at $t_f$ is not simply the sum of time-dependent impacts from $t_0$ to $t_f$ due to different GHG lifetimes and more complex behaviors that are currently simplified with the use of a GWP100 LCIA method. While this mismatch may be limited for climate change, it illustrates remaining inconsistencies that should be addressed. Future work could also explore the implications of time-explicit LCA for temporal dynamics in allocation~\citep{muller2025time}, and integrate interactions between carrying capacities, which are increasingly discussed in the literature~\citep{lade2020human, lejeune2026pathways}.

\subsection{Perspectives and recommendations for adoption in the~community}

We argue that supporting the development of AESA requires a minimum level of harmonization in implementation. Implementing the proposed framework from scratch would require considerable time, expertise, and resources, which would limit its uptake at scale. For this reason, we provide \texttt{pyaesa}, the open-source Python implementation of \textit{UNCASExt}, together with extensive supplementary material to support transparency and reproducibility (see Appendices).

The proposed framework and its Python implementation provide a practical basis for the AESA community to experiment, compare, and incrementally refine AESA practices toward greater harmonization. In fact, they ensure that the different sharing principles considered are consistent with the defined scope and demand perimeter, and that the uncertainties associated with each enacting metric and allocation path are properly propagated to the output, allowing the main drivers of output variability to be identified. However, practitioners remain responsible for deciding whether (i) all available sharing principles are considered, as recommended in the default mode of \texttt{pyaesa}, or (ii) only a subset is selected, in which case it must be clearly justified in the study's goal and scope. To further foster harmonization and transparency in the AESA community, a shared repository is also available on GitHub (\url{https://github.com/AESAtoolkit/AESA_case_studies}) for practitioners to share implementations of their case studies developed with \texttt{pyaesa}. 

\textit{UNCASExt} and \texttt{pyaesa} can also be used to explore new allocation methods (e.g., based on the Human Development Index as proposed by~\citep{verhaeghe2024carrying}), thereby further supporting the investigation of different distributive justice theories. This work can also provide a strong basis for allocating carrying capacities at lower levels, e.g., company~($L_3$) and product~($L_4$) levels (see \appendixEM~for L3 and L4 functional units). For AESA at the product level, a ``product'' should be understood in the \textit{UNCASExt} framework as a ``good or service'', similarly to LCA ISO standards. The product level is therefore not restricted to outputs supplied to final demand: it can also include intermediate products, e.g., components, processes, etc. This is an important methodological contribution of the \textit{UNCASExt} framework, as it covers not only B2C but also B2B activities.
Nevertheless, allocation at these lower levels still requires (i)~further methodological development in terms of specific equations to cover the wide range of rapidly evolving product functionalities, and (ii)~highly case-specific data. It is therefore much more difficult, or even impossible, to implement systematically using currently available MRIO databases alone. External data sources can be used, but this increases data heterogeneity and the risk of scope mismatch. In this context, the release of an enterprise-level MRIO database~\citep{kata2025} could serve as a promising data source for future, more systematic lower-level AESA studies. More generally, there is currently no clear consensus in the AESA community regarding the extent to which it is relevant to conduct AESA at lower levels, which further highlights the need for future work.

\section{Conclusions} \label{sec:conclusion}

In recent years, AESA has gained increasing attention within the environmental assessment research community, and its application has progressively expanded beyond academia to inform policy-making. However, the wide range of uncertainty sources encountered during an AESA study is often poorly accounted for or overlooked, result variability remains difficult to attribute to individual uncertainty sources, and scope inconsistencies between estimated environmental burdens and allocated carrying capacities are not systematically addressed. Together, these limitations undermine confidence in the interpretation and robustness of AESA results. \par

The \textit{UNCASExt} framework introduced in this work directly addresses these gaps. 
First, \textit{UNCASExt} broadens the set of uncertainty sources covered from Phase~A to Phase~C and enables the first statistical quantification in AESA of how individual uncertainty sources contribute to result variability. \par 

Second, the framework strengthens scope consistency between estimated environmental burdens and allocated carrying capacities by providing 21~combinations of system boundaries and functional units, 28~corresponding enacting metrics, 36~individual allocation equations, and 84~allocation paths. In fact, this work demonstrates quantitatively that scope mismatch within the ASR is a critical source of relevance uncertainty in AESA, even though it remains insufficiently addressed in the existing literature. \par 

Third, we provide an open-source Python package, \texttt{pyaesa}, alongside the \textit{UNCASExt} framework, enabling reproducible and systematic implementation, as well as the allocation of carrying capacities from the global level to countries and sectors for all functional units introduced in this work. \par

The framework also incorporates temporal dynamics by supporting both retrospective and prospective AESA with either static steady-state or dynamic carrying capacities. This includes defining dynamic GHG budgets as carrying capacities to explore transition pathways. \par

The practical implementation of the framework is illustrated through a case study assessing the absolute environmental sustainability of France's electricity consumption from 2019 to 2060. By evaluating prospective scenarios for the future French electricity mix, the case study demonstrates how \textit{UNCASExt} and \texttt{pyaesa} can support decision-making by comparing transition pathways against allocated carrying capacities. The results show that although the assessed scenarios improve absolute sustainability performance over time, they are insufficient to ensure that France's electricity consumption remains within all allocated carrying capacities by 2060. The case study also demonstrates how the framework supports result interpretation by attributing result variability to individual uncertainty sources. The choice of allocation method remains the main driver across impact categories, although its contribution decreases in the prospective assessment as temporal and scenario uncertainties become more influential, from 95\%--99\% in 2019 to 64\%--97\% in 2060.\par

Beyond this case study, \textit{UNCASExt} and the open-source Python package \texttt{pyaesa} provide a scalable, reproducible, and transparent computational AESA framework. Designed for LCA and AESA practitioners, the framework aims to accelerate the adoption of robust sustainability assessments to inform decision-making within planetary boundaries. By bridging methodological guidelines and real-world implementation through \texttt{pyaesa}, the framework offers a flexible yet harmonized foundation for AESA studies. The open-source nature of this work, strengthened by comprehensive appendices, ensures accessibility and reproducibility, thereby supporting broader uptake and collaborative refinement by the community.\par

\section{Supporting Information} \label{sec:appendixes}

The full \textit{UNCASExt} framework is provided as an open-source Python package \texttt{pyaesa} on GitHub at~\url{https://github.com/AESAtoolkit/pyaesa} under a GPL-3.0 license. It can be downloaded from \texttt{PyPI} and \texttt{conda-forge}. Extensive documentation is provided with the package on its repository and via readthedocs.io at~\url{https://pyaesa.readthedocs.io/en/latest/}. \par

The code to generate the results for the case study based on \texttt{pyaesa} is provided at~\url{https://github.com/AESAtoolkit/AESA_case_studies} in the \textit{level\_2\_sector} folder. \par 

Moreover, several appendices are provided as supplementary information to ensure full transparency and reproducibility of the framework. This aims at supporting adoption by a wide range of practitioners in the AESA community. \par 

\textbf{\appendixComparison} provides a fully transparent account of the scientific advances with respect to the original UNCASE framework. It details the key features and limitations of the original UNCASE framework~\citep{puig2025quantifying} and provides an extensive comparison with the proposed framework. \par

\textbf{\appendixEM} details all allocation paths, accounting system boundaries, and functional units covered by \textit{UNCASExt} at each allocation level. Equations are provided for the 28~enacting metrics and 36~individual allocation methods as well as the validation procedure for the \texttt{pyaesa} implementation. Finally, data and uncertainty sources are also presented.
\par

\textbf{\appendixCC} details carrying capacity definitions, both for static and dynamic approaches. Values are provided for global static CCs~\citep{kitzmann2025planetary, bjorn2015introducing, sala2020environmental, sanye2023consumption}. The method developed to systematically define alternative dynamic carrying capacities for climate change is detailed.

\textbf{\appendixCaseStudy} provides additional information regarding the case study modeling, results, and limitations. \par

\section*{Acknowledgments}

The authors would like to thank Anders Bj$\o$rn, Michaël Lejeune,  Romain Sacchi, and Joanna Schlesinger-Martinat for the valuable discussions and interactions during the realization of this work. The authors also thank Niklas von der Assen, Mikolaj Owsianiak, Natacha Gondran, Timo Diepers, Sankalp Shrivastava, Andrea Paulillo and Julie Clavreul for discussions in the context of the Python package \texttt{pyaesa}.

\section*{CRediT authorship contribution statement}

\textbf{Erwan Ike de Bantel:} Writing – original draft, Visualization, Software (lead developer), Methodology, Investigation, Formal analysis, Conceptualization; Writing – review \& editing, Validation, Conceptualization.
\textbf{Thibault Pirson:} Writing – original draft, Visualization, Software, Methodology, Investigation, Formal analysis, Conceptualization; Writing – review \& editing, Validation, Conceptualization. 
\textbf{Gonzalo Puig-Samper:} Writing – original draft, Writing – review \& editing, Conceptualization, Software. 
\textbf{Jan Hartmann:} Writing - Review \& editing, Conceptualization, Software. 
\textbf{David Bol:} Conceptualization, Writing - Review \& editing, Resources, Supervision, Funding acquisition.
\textbf{Ghada Bouillass:} Conceptualization, Writing - Review \& editing, Resources, Supervision.
\textbf{Bernard Yannou:} Conceptualization, Writing - Review \& editing, Resources, Supervision, Funding acquisition.
\textbf{Marija Jankovic:} Conceptualization, Writing - Review \& editing, Resources, Supervision.
\textbf{Michael Hauschild:} Conceptualization, Writing - Review \& editing, Resources, Supervision.
\\

\textbf{Generative AI} tools were used during the development of \texttt{pyaesa}. As the development of the package took several months from the initial ideation to the first release, several tools were used including OpenAI Codex models 5.3 to 5.5, Codex Spark 5.3, Anthropic Claude Opus 4.6 and Sonnet 4.6, and GitHub Copilot.
These tools were used to support implementation tasks, including code generation, refactoring, debugging, test design, documentation drafting, code review, computing time optimization, memory use reduction, and figure rendering workflows.
The development of AESA methodology and workflow, mathematical expressions, software architecture and the design of the figures were defined by the authors. Generative AI outputs were used as implementation drafts, checked and modified where needed, and retained only after review and testing by human intelligence.

\section*{Funding Information}

The work of Erwan Ike de Bantel received organizational support of the "CircularIT Alliance" project.
The work of Thibault Pirson is part of the SOIL project, which received funding from the European Union’s Horizon Europe research and innovation program under the HORIZON-KDT-JU-2023-1-IA grant agreement N°101139785. Views and opinions expressed are however those of the authors only and do not necessarily reflect those of the European Union or CHIPS. Neither the European Union nor the granting authority can be held responsible for them.

\section*{Data Availability Statement}

All information regarding data sources can be found directly in the documentation of \texttt{pyaesa} on GitHub at~\url{https://github.com/AESAtoolkit/pyaesa} or in the readthedocs.io at~\url{https://pyaesa.readthedocs.io/en/latest/}. 

Data for the case study can be retrieved on GitHub at \url{https://github.com/oie-mines-paristech/RTE_scenarios}. Case study results can be fully reproduced via two dedicated notebooks available at~\url{https://github.com/AESAtoolkit/AESA_case_studies}, which further demonstrates the added value of using \texttt{pyaesa} to ensure reproducibility of AESA studies at scale.

\section*{Declaration of competing interest}

The authors declare that they have no known competing financial interests or personal relationships that could have appeared to influence the work reported in this paper.

\bibliographystyle{elsarticle-harv}
\bibliography{bibliography.bib}{}

\end{document}